\pdfoutput=1 

\documentclass[review, 12pt]{elsarticle}

\usepackage{amssymb}
\usepackage{bm}
\usepackage{amsmath}
\usepackage{caption}
\usepackage{cleveref}
\usepackage{multirow}
\usepackage{color,soul}
\usepackage[table]{xcolor}
\biboptions{compress}

\begin{document}
\sethlcolor{yellow}
\setstcolor{red}
\soulregister\cite7 
\soulregister\ref7

\begin{frontmatter}

    \title{A fully conservative sharp-interface method for compressible mulitphase flows with phase change}

    \author{Tian Long}

    \author{Jinsheng Cai}

    \author{Shucheng Pan\corref{cor1}}
    \ead{shucheng.pan@nwpu.edu.cn}

    \cortext[cor1]{Corresponding author}

    \address{School of Aeronautics, Northwestern Polytechnical University, Xi'an, 710072, PR China}

    \begin{abstract}
    A fully conservative sharp-interface method is developed for multiphase flows with phase change. The coupling between two phases is implemented via introducing the interfacial fluxes, which are obtained by solving a general Riemann problem with phase change. A novel four-wave model is proposed to obtain an approximate Riemann solution, which simplifies the eight-dimensional roo-finding procedure in the exact solver to a sole iteration of the mass flux. Unlike in the previous research, the jump conditions of all waves are imposed strictly in the present approximate Riemann solver so that conservation is guaranteed. Different choices of the fluid states used in the phase change model are compared, and we have shown that the adjacent states of phase interface should be used to ensure numerical consistency. To the authors' knowledge, it has not been reported before in the open literature. With good agreements, various numerical examples are considered to validate the present method by comparing the results against the exact solutions or the previous simulations. 

    \end{abstract}

    \begin{keyword}
        Sharp-interface method \sep Fully conservative \sep Multiphase flows \sep Phase change  
    \end{keyword}

\end{frontmatter}


\section{Introduction}
\label{sec1}
The multiphase flows with phase change are ubiquitous in scientific and industrial applications such as cavitation in the naval industry \cite{Bonfiglio2016multiphase}, bubble condensation in electronic cooling \cite{liu2020assessment}, the combustion of fuel in rocket engines \cite{mayer1996propellant}, etc. Over the last decades, many numerical methods have been developed to investigate the underlying mechanisms of these flow phenomena, which can be generally divided into two classes, i.e. diffuse interface methods and sharp interface methods. In the diffuse interface methods, the material interface is smeared out over a finite number of grid cells where different fluids are mixed \cite{ding2007diffuse}. For the mixed fluids at the interface region, a mixture model or an artificial equation of state (EOS) needs to be developed to obtain a thermodynamically consistent description \cite{allaire2002fiveequation}, which is difficult as the mixture states may be even unphysical. A typical example is the homogeneous equilibrium method (HEM) proposed by Menikoff and Plohr \cite{menikoff1989riemann}, in which the mixed fluids were assumed to be a mixture of liquid and vapor in thermodynamic phase equilibrium. This approach was further extended to the van der Waals EOS by M{\"u}ller and Vo{\ss} \cite{muller2006riemann} via using the Maxwell construction. However, the phase equilibrium assumption suffers a kink at the coexistence curves of two phases \cite{hitz2021comparison}. The seven-equation model \cite{baer1986twophasemixture,embid1992mathematical,saurel1999multiphase} is a full non-equilibrium model, in which instantaneous relaxation procedures are imposed for pressure and velocity. By including temperature and chemical potential relaxation effects, Zein et al. \cite{zein2010modeling} modified this model to take into account phase transition. Nevertheless, the solutions of the seven-equation model usually exhibit a complex wave structure \cite{kapila2000twophase}. In addition, in the diffuse interface methods, the interface profile tends to be thickened over time due to numerical diffusion. Although several interface sharpening schemes have been developed to suppress the interface diffusion \cite{kokh2010anti,shukla2010interface,so2012anti}, it remains a big challenge to conduct long-time simulations with the diffuse interface methods \cite{lin2017simulation}.

In the sharp-interface methods, the interface is modeled as a discontinuity within the flow filed \cite{glimm2002interface,glimm2003conservative,chang2013direct}, at which appropriate jump conditions need to be imposed to ensure the physical consistency of the overall model. To track the interface with a non-smearing representation, different approaches have been developed, e.g., the front-tracking method \cite{unverdi1992front,tryggvason2001front}, where the interface is represented by massless markers, the arbitrary Lagrangian Eulerian (ALE) method \cite{Hirt1974arbitrary,Ling2010numerical}, where the interface coincides with mesh lines, and the level-set method \cite{osher1988fronts}, where the interface is defined as the zero level-set of a signed distance function. When phase change is considered, although there have been many efforts in the context of incompressible multiphase flows (see Refs. \cite{gibou2007phasechange,sato2013phasechange,tanguy2014phasechange,shaikh2016stefan,lee2017phasechange}, to name but a few), only a few numerical schemes for compressible regimes are reported in the open literature and all of them are based on the level-set method. In contrast to other approaches, the level-set method is simple to implement and can handle large interface deformations and topological changes in an automatical way \cite{osher2001level}, which make it more suitable for modelling phase change in compressible multiphase flows. In Ref. \cite{lee2017sharp}, the level-set method was combined with a ghost fluid method (GFM) \cite{fedkiw1999ghost} to investigate the compressible bubble growth with phase change. The states of ghost cells, i.e., the cells defined in the vicinity of the interface where the ghost fluid and real fluid co-exist, were determined by the jump conditions accounting for phase transfer. However, as shown by Liu et al. \cite{liu2003ghost}, the ghost cells should be populated according to the solution of an interfacial Riemann solver so that waves can be correctly transmitted and reflected at the interface. Following this strategy, Houim and Kuo \cite{houim2013ghost} extended the GFM to reacting flows with phase change by solving a modified interfacial Riemann problem. The resulting velocity and pressure jumps due to phase change were incorporated in the conservation equations of the gas-liquid interface. This work was further modified by Das and Udaykumar \cite{das2020sharp} via simplifying the implementation of the interfacial jump conditions. The numerical instabilities caused by the rotation of the deviatoric stress tensor were avoided by rotating the velocity field. Since the velocity and pressure jumps are calculated using the left and right states of the interfacial Riemann problem, one common issue in Refs. \cite{houim2013ghost,das2020sharp} is that the conservation equations are not truly fulfilled at the interface. By introducing an additional wave representing a phase interface, which was observed in several experiments \cite{kurschat1992complete,simoes1999evaporation,reinke2001explosive}, Fechter et al. proposed an exact Riemann solver \cite{fechter2017sharp} and an approximate solver \cite{fechter2018approximate} for a general two-phase Riemann problem including phase change and surface tension. Note that, in Refs. \cite{fechter2017sharp,fechter2018approximate}, the coupling between two phases at the interface is also imposed through the GFM.

However, for all the GFM-based methods, an unavoidable problem is the lack of conservation properties, which is crucial for compressible multiphase flows with strong interface interactions. For compressible multiphase flows without phase change, a fully conservative sharp-interface method was proposed by Hu et al. \cite{hu2006conservative}. This method was extended by Lauer et al. \cite{lauer2012numerical} to simulate oscillating bubbles with phase change. They assumed that phase change is much slower than the hydrodynamic interface interaction, which may not hold for other cases. A more general model taking account of evaporation is independently developed by Paula et al. \cite{paula2019analysis}. Since the heat flux at the phase boundary is neglected in Ref. \cite{paula2019analysis}, this model will lead to inconsistent stationary solutions, see Ref. \cite{fechter2017sharp} for a detailed derivation.

In this paper, we aim to develop a fully conservative sharp-interface method based on \cite{hu2006conservative} for compressible multiphase flows with phase change. Unlike in Refs. \cite{lauer2012numerical,paula2019analysis}, we obtain interfacial fluxes through solving the general Riemann problem proposed by Fechter et al. \cite{fechter2017sharp} to ensure thermodynamic consistency. The exact Riemann solver proposed in Ref. \cite{fechter2017sharp} suffers from poor efficiency due to the resulting eight-dimensional root-finding problem. To address this issue, a novel approximate Riemann solver is developed in this paper. Unlike the approximate Riemann solver of fechter et al. \cite{fechter2018approximate}, the present solver employs a four-wave model so that the energy coupling in the vapor phase is achieved. Since the specific form of an EOS is not used in this Riemann solver, it can be used to simulate real fluids. In addition, the choices of the liquid and vapor states used in the phase change model is investigated in detail, which is not discussed in the previous researches. As we will show in Section. 4.1, the intermediate states in the Riemann solution fan should be used to calculate the mass flux to ensure numerical consistency. The remainder of this paper is organized as follows.
In Section 2, we briefly review the conservative sharp-interface framework for multiphase flows without phase change. Subsequently in Section 3 we introduce the novel approximate Riemann solver for the general Riemann problem including phase change and surface tension effects. A number of numerical
tests are carried out to validate the present method in Section 4, followed by the concluding remarks in Section 5.

\section{Conservative sharp-interface framework}
\label{sec2}

\subsection{Governing equations}
For inviscid compressible flows, the governing equations are
\begin{equation}
    \frac{\partial \bm{U}}{\partial t}+\nabla \cdot \bm{F(U)} = 0 ,
    \label{Eq:NSequations}
\end{equation}
where $\bm{U} = (\rho, \rho u, \rho v, \rho w, \rho E)^T$ is the vector of conserved variables, in which $\rho$, $u$, $v$, $w$ and $E$ denote the density, the three Cartesian velocity components and the total energy with relation $E = e + \frac{1}{2}(u^2 +v^2 +w^2)$, respectively. Note that $e$ stands for the internal energy per unit mass. The inviscid flux tensor $\bm{F}$ reads
\begin{equation}
    \begin{aligned}
         & \bm{F(U)} =\begin{bmatrix}
            \rho u        & \rho v        & \rho w        \\
            \rho u^2 + p  & \rho v u      & \rho w u      \\
            \rho u v      & \rho v^2 + p  & \rho w v      \\
            \rho u w      & \rho v w      & \rho w^2 + p  \\
            u(\rho E + p) & v(\rho E + p) & w(\rho E + p)
        \end{bmatrix}, \\ \\
    \end{aligned}
    \label{Eq:inviscid flux}
\end{equation}
and $p$ is the pressure. To close this system, an EOS is needed to describe the thermodynamic properties of the materials. Thanks to the sharp-interface treatment employed in this paper, different EOS can be used in the bulk region of each phase so that any unphysical mixing is avoided. The specific thermodynamic description is given in the following section.
\subsection{Two-phase thermodynamics}
With an evolving interface $\Gamma (t)$, the flow domain $\Omega$ is divided into two sub-domains $\Omega_{liq}$ and $\Omega_{vap}$, which represent the regions occupied by the liquid phase and the vapor phase, respectively. Assuming that the fluid in each region is in a local thermodynamic equilibrium, its state can be determined by any two independent thermodynamic variables. In this paper, the fluids water and n-dodecane are considered, which are described by the stiffened-gas EOS \cite{paula2019analysis,zein2010modeling} and the Helmholtz-energy-based EOS \cite{lemmon2004thermodynamic}, respectively. 

For the stiffened-gas EOS, the relations between different thermodynamic variables are given by
\begin{equation}
    \begin{aligned}
         e(p,\rho) &= \frac{p +\gamma p_{\infty}}{(\gamma - 1)\rho} + e_{ref},\\
         T(p,\rho) &= \frac{p + p_{\infty}}{C_v(\gamma - 1)\rho},
    \end{aligned}
    \label{Eq:stiffened gas EOS}
\end{equation}
where $T$ is the temperature, $C_v$ the heat capacity at constant volume, $e_{ref}$ the reference internal energy, $\gamma$  the adiabatic coefficient and $p_{\infty}$ the parameter accounting for the pre-compression of the fluid. Following Ref. \cite{zein2010modeling}, we use $\gamma = 2.35$, $p_{\infty}=10^9\ \rm{Pa}$, $e_{ref} = -1167\times10^3\ \rm{J / kg}$ and $C_v = 1.816\times10^3\ \rm{J/kg/K}$ for liquid water while in the vapor phase, we use $\gamma = 1.33$, $p_{\infty}=0.0\ \rm{Pa}$, $e_{ref} = 1990\times10^3\ \rm{J / kg}$ and $C_v = 1.399\times10^3\ \rm{J/kg/K}$ so that Eq. \eqref{Eq:stiffened gas EOS} degenerate to the ideal-gas EOS \cite{paula2019analysis}. 

In the Helmholtz-energy-based EOS, a non-analytical formulation of the residual Helmholtz energy is employed, whose parameters are obtained by fitting the underlying EOS to the experimental data. Although it is very inefficient, by using this kind of EOS, the prediction error for fluid properties is usually less than $\%1$ \cite{lemmon2006short}. Moreover, the terms used in this kind of EOS varies for different materials so that we cannot simulate different fluids by simply changing the parameters. Here n-dodecane is considered to validate the present method for real-fluid simulations. The specific formulation is not given here for simplicity and the interested readers are referred to Ref. \cite{lemmon2004thermodynamic}.

As phase change is considered, only the fluid states below the critical point \cite{fechter2017sharp} are considered such that the liquid phase and the vapor phase co-exist. We also require that the so-called saturation curves exist, which can be characterized by 
\begin{equation}
    \begin{aligned}
    T_{liq}^{sat} &= T_{vap}^{sat},\\
    G_{liq}^{sat} &= G_{vap}^{sat}, \\
    p_{liq}^{sat} &= p_{vap}^{sat} +\sigma \kappa,
    \end{aligned}
\end{equation}
where $G$ is the Gibbs free energy, $\sigma$ the surface tension coefficient, and $\kappa$ the interface curvature.
As in Ref. \cite{fechter2018approximate}, we restrict ourselves to admissible EOS regions, where, with $S$ being the entropy, the fundamental derivative for sound speed $\left. \frac{\partial p}{\partial \rho}  \right|_S$ is positive. By doing so, the split and composite waves \cite{hitz2021comparison,fechter2017sharp} are excluded from the two phase Riemann problem considered in Section. \ref{sec3}, whose treatment is beyond the scope of the present study. 

\begin{figure}[htbp]
    \centering
    \includegraphics[width=0.6\textwidth]{ 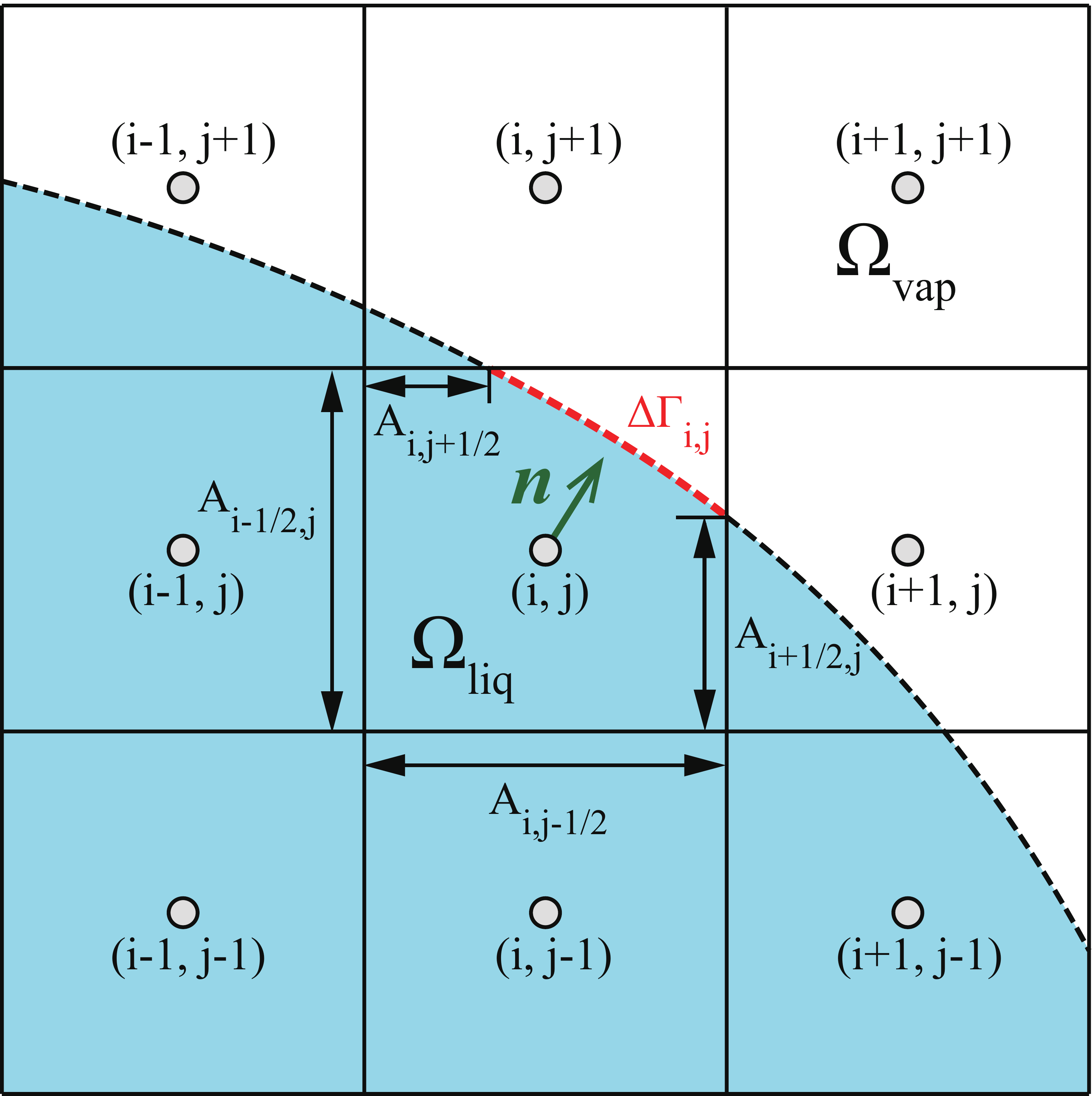}
    \caption{Two-dimensional schematic of the conservative discretization in a cut cell $(i, j)$. The blue domain $\Omega_{liq}$ and white domain $\Omega_{vap}  $ are occupied by the liquid phase and the vapor phase, respectively. The red dashed line and green arrow indicates the interface segment $\Delta \Gamma_{i,j}$ in this cell and the normal vector at the cell center, respectively.}
    \label{Fig:conservation_discretization}
\end{figure}

\subsection{Conservative discretization}
The main principle of the conservative sharp-interface method \cite{hu2006conservative} is to modify the spatial discretization in the Cartesian cells cut by the interface, i.e., the so-called cut cells. Take a two-dimensional Cartesian grid with spacing $\Delta x$ and $\Delta y$ as an example. By applying the first-order forward Euler method to Eq. \eqref{Eq:NSequations} leads to the following discretized equations,
\begin{equation}
    \begin{aligned}
    {\alpha}^{n+1}_{i,j}{\bm{U}}^{n+1}_{i,j}=&{\alpha}^{n}_{i,j}{\bm{U}}^{n}_{i,j}+\frac{\Delta t}{\Delta x \Delta y}\hat{\bm{X}}(\Delta {\Gamma}_{i,j}) +\frac{\Delta t}{\Delta x}[{A}_{i-1/2,j}{\hat{\bm{F}}}_{i-1/2,j}-{A}_{i+1/2,j}{\hat{\bm{F}}}_{i+1/2,j}] \\
    &+\frac{\Delta t}{\Delta y}[{A}_{i, j-1/2}{\hat{\bm{F}}}_{i, j-1/2}-{A}_{i, j+1/2}{\hat{\bm{F}}}_{i, j+1/2}],\\
    \end{aligned}
    \label{Eq:NSdiscrete}
\end{equation}
where $\Delta t$ is the time step and $\alpha_{i,j} \bm{U}_{i,j}$ is the conservative variables in the cut cell, with $\alpha_{i,j}$ being the volume fraction of the corresponding phase. As illustrated in Fig. \ref{Fig:conservation_discretization}, $A$ and $\Delta \Gamma$ represent the cell-face apertures and the interface segment of the cut cell, respectively. The reconstructed cell-face numerical inviscid flux $\hat{\bm{F}}$ is obtained by a high-order shock-capturing scheme such as the WENO \cite{shu1988efficient} scheme. The interfacial flux $\hat{\bm{X}}$, which is crucial for modelling phase change, is obtained by solving a two-phase Riemann problem along the normal direction $\bm{n}$ (see green arrow in Fig. \ref{Fig:conservation_discretization}). Let $\bm{W}^{l}_{i,j}(\rho,V,p,e)$ and $\bm{W}^{v}_{i,j}$ denote the primitive variables of the liquid phase and the vapor phase in the cut cell (i, j), respectively, where $V = \bm{u}\cdot\bm{n}$ is the normal velocity. Then the two-phase Riemann problem can be formulated as
\begin{equation}
    \mathcal{R}_{i,j}=\left\{
    \begin{aligned}
         & \mathcal{R}(W^l_{i,j}, W_{i, j}^{v, (g)}),  \quad \text{if the cell center $(i,j)$ is located in $\Omega_{liq}$}, \\
         & \mathcal{R}(W^{l,(g)}_{i,j}, W_{i, j}^{v}),  \quad \quad \text{otherwise},
    \end{aligned}
    \right.
    \label{interfacial_Riemann_problem}
\end{equation}
in which the superscript (g) indicates the ghost states obtained by using the extending algorithm \cite{fu2017single}. 

\subsection{Interface description}
The interface is tracked by a level set function $\phi(\bm{x},t)$, which represents the signed distance from the interface to each cell center. The time evolution of $\phi(\bm{x},t)$ \cite{osher1988fronts} is governed by
\begin{equation}
    \frac{\partial \phi}{\partial t}+\hat{\bm{u}} \cdot \nabla \phi = 0,
    \label{LS:linearadvection}
\end{equation}
where $\hat{\bm{u}}$ represents the interface velocity obtained by solving Eq. \eqref{interfacial_Riemann_problem}. The normal direction and the curvature can be calculated by
\begin{equation}
    \bm{n} = \nabla \phi \quad \text{and} \quad \kappa = \nabla \cdot \frac{\nabla \phi}{|\nabla \phi|},
\end{equation}
respectively.

\section{Approximate Riemann solution with phase change}
\label{sec3}

\subsection{The general Riemann problem with phase change}
\label{sec3.1}
\begin{figure}[htbp]
    \centering
    \includegraphics[width=1.0\textwidth]{ 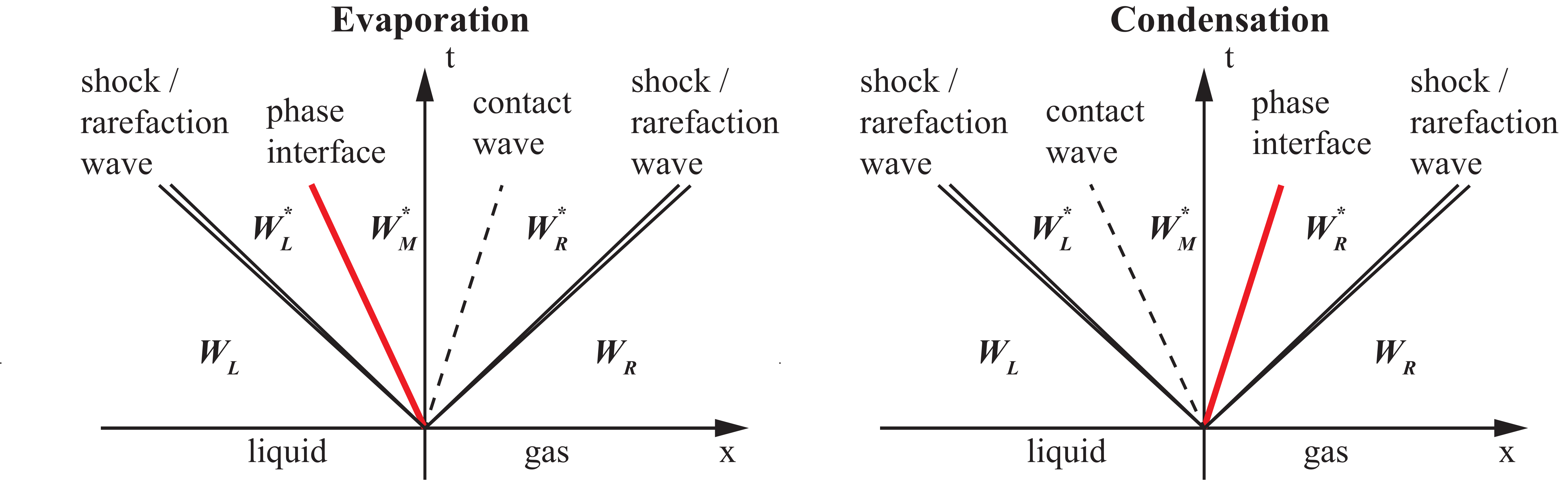}
    \caption{The wave structures for the general Riemann problem with evaporation and condensation.}
    \label{Fig:exact_riemann}
\end{figure}
As mentioned above, a one-dimensional two-phase Riemann problem is solved along the normal direction to obtain the interfacial flux. When phase change is considered, an additional wave is introduced in the Riemann solution fan to model the phase interface. Assuming a left liquid state and a right vapor state, we plot the Riemann wave patterns for evaporation and condensation in Fig. \ref{Fig:exact_riemann} with a $(x,t)$-diagram. For the other three waves, standard jump conditions \cite{toro2013riemann} can be imposed directly. To ensure thermodynamic consistency \cite{fechter2017sharp}, the jump conditions at the phase interface are given by  
\begin{equation}
    \begin{aligned}
    \left[\rho(V - S_p)\right] & = 0,\\
    \left[\rho(V - S_p)V + p \right] & = -\sigma \kappa, \\
    \left[\rho(V - S_p)(e +\frac{1}{2}V^2) + p V \right] & = j Q_{lat} - \sigma \kappa S_p,
    \end{aligned}
    \label{eq:jump_conditions_pc}
\end{equation}
where $[a] = a^*_{vap} - a^*_{liq}$ is the jump operator, $Q_{lat}$ the latent heat set as $2240000\ \rm{J/kg}$ for water and $249410\ \rm{J/kg}$ for n-dodecane, $\sigma$ the surface tension coefficient, $S_p$ the velocity of the phase interface and $j$ the mass flux with relation
\begin{equation}
  j = \rho^*_{liq}(V^*_{liq} - S_p)= \rho^*_{vap}(V^*_{vap} - S_p).
  \label{eq:mass_flux}
\end{equation}
Here the superscript $*$ denotes the states in the intermediate region, the so-called star region, and the subscripts $liq$ and $vap$ are chosen according to the direction of phase change, which is formulated as
\begin{equation}
    (liq,vap) = \left\{
    \begin{aligned}
        &(L,M) \quad \text{if $j > 0$ (evaporation),}\\
        &(M,R) \quad \text{if $j < 0$ (condensation).}\\
    \end{aligned}
    \right.
    \label{subscripts}
\end{equation}
As in Refs. \cite{fechter2017sharp,fechter2018approximate}, the phase interface is assumed to be subsonic and non-characteristic,
\begin{equation}
    \left\{
    \begin{aligned}
    V^*_{liq} - c^*_{liq} &< S_p < V^*_{liq} + c^*_{liq},\\
    V^*_{vap} - c^*_{vap} &< S_p < V^*_{vap} + c^*_{vap},
    \end{aligned}
    \right.
\end{equation}
in which $c$ is the sound speed. With the phase interface, the resulting system is underdetermined and an additional phase change model is needed to decide the mass flux $j$.

\subsection{The four-wave approximate Riemann solver}
\begin{figure}[htbp]
    \centering
    \includegraphics[width=1.0\textwidth]{ 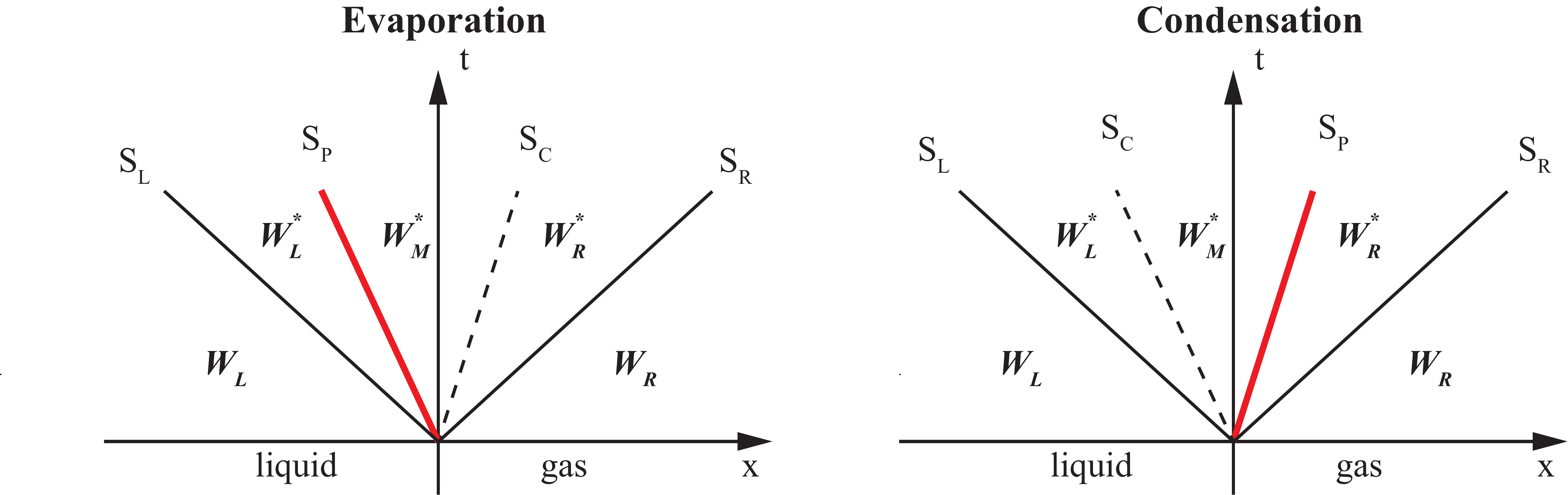}
    \caption{The wave structures for the approximate Riemann solution with evaporation and condensation.}
    \label{Fig:approximate_riemann}
\end{figure}
Based on the work of Fechter et al. \cite{fechter2018approximate}, a novel approximate Riemann solver is developed by reconstructing a contact wave in the Riemann wave fan. As illustrated in Fig. \ref{Fig:approximate_riemann}, the solution consists of five regions with constant states, which are separated by four waves including two shock waves, a phase interface, and a contact wave. In the approximate solution, the wave speeds of the shock waves are estimated a priori. In this paper, we use the simple estimates suggested by Davis \cite{davis1988simplified}, which read
\begin{equation}
    S_L = V_L - c_L \quad \text{and}\quad S_R = V_R + c_R.
    \label{eq:wave_speeds}
\end{equation}
The jump conditions at the shock waves are given by
\begin{equation}
    \begin{aligned}
        \rho_k^*(V_k^* - S_k)                                        & = \rho_k(V_k - S_k),                              \\
        \rho_k^*(V_k^* - S_k)V_k^* + p_k^*                           & = \rho_k(V_k - S_k)V_k + p_k,                     \\
        \rho_k^*(V_k^* - S_k) (e^*_k + \frac{1}{2}{V_k^*}^2) + p_k^* & = \rho_k(V_k - S_k)(e_k + \frac{1}{2}V_k^2) +p_k,
    \end{aligned}
    \label{eq:jump_conditions_shock}
\end{equation}
where $k=L,R$. For the contact wave, the standard Rankine-Hugoniot equations can be simplified to 
\begin{equation}
    \begin{aligned}
    V_M^* &= S_c,\quad V_M^* &= V_k^*,\quad p_M^* &= p_k^*,
    \label{eq:jump_conditions_contact}
    \end{aligned}
\end{equation}
in which $S_c$ denote the speed of the contact wave, $k = L$ or $k = R$ represents the condensation case or the evaporation case, respectively. Up to now, with the jump conditions of the four waves Eqs. (\ref{eq:jump_conditions_pc}, \ref{eq:jump_conditions_shock}, \ref{eq:jump_conditions_contact}) and the relation of the mass flux Eq. \eqref{eq:mass_flux}, we obtain a simplified system with thirteen independent equations. However, there are fifteen known variables in total: $\bm{W_L}^*(\rho_L^*,V_L^*,p_L^*,e_L^*)$, $\bm{W_M}^*$, $\bm{W_R}^*$, $S_p$, $S_c$, and $j$. Note that the pressure in the star region does not agree with the thermodynamic pressure calculated by using the EOS, which is dropped due to the HLL approximation \cite{harten1983upstream}. 

To solve the current system, two additional equations need to be added. For the first one, in this paper, we make an assumption that
\begin{equation}
    \frac{\rho^*_{liq}}{\rho^*_{vap}}  \approx  \frac{\rho_L}{\rho_R}.
    \label{Eq:density_assumption}
\end{equation}
The second equation is a phase change model for calculating the mass flux,
\begin{equation}
    j = f_m(\bm{W_{liq}}^*,\bm{W_{vap}}^*).
    \label{Eq:phase_change_model}
\end{equation}
Following Refs. \cite{houim2013ghost,das2020sharp}, we employ the Hertz-Knudsen relation (HKR) \cite{schrage1953theoretical},
\begin{equation}
    f_j = \frac{1}{2\pi R_v}\left(\lambda_{evap}\frac{p_{sat}(T^*_{liq})}{\sqrt{T^*_{liq}}} - \lambda_{cond}\frac{p^*_{vap}}{\sqrt{T^*_{vap}}} \right),
    \label{Eq:HKR}
\end{equation}
where $p_{sat}$ is the saturation pressure, $R_v$ is the specific gas constant, $\lambda_{evap}$ and $\lambda_{cond}$ are the evaporation and condensation coefficients, respectively. In this paper, $R_v$ is set to $461.52\ \rm{J/(kg\cdot K)}$ for water and $48.81\ \rm{J/(kg\cdot K)}$ for n-dodecane. As in Ref. \cite{persad2016expressions}, the saturation pressure of water is calculated by 
\begin{equation}
\begin{aligned}
 p_{sat}(T) =\ &611.2\ \text{exp}(1045.8511577 - 21394.6662629 T^{-1} + 1.0969044 T \\
              &- 1.3003741 \times 10^{-3} T^{2} + 7.7472984 \times 10^{- 7} T^3 \\
              &- 2.1649005 \times 10^{-12} T^4 - 211.3896559  \ln{T} ),
\end{aligned}
\label{Eq:saturation_pressure_water}
\end{equation}
while the saturation pressure of n-dodecane is obtained via the open-source library CoolProp \cite{bell2014pure}. Special attention needs to be paid to the calculation of the temperature. As mentioned above, in the star region, the density, the pressure, and the internal energy do not agree with the EOS. In consequence, we have three ways to calculate the temperature: $T^* = T^*(\rho^*,p^*)$, $T^* = T^*(\rho^*,e^*)$, and $T^* = T^*(p^*,e^*)$. The last one is used in the present method to eliminate the approximate error of density introduced by Eq. \eqref{Eq:density_assumption}.

By using all equations except the phase change model Eq. \eqref{Eq:phase_change_model}, the unknown variables $\bm{W_M}^*$, $\bm{W_R}^*$, $S_p$, and $S_c$ can be expressed in terms of $j$, which are given in detail in Appendix A. The one-dimensional reduced system can be expressed by 
\begin{equation}
    (\bm{W_{L}}^*,\bm{W_{M}}^*,\bm{W_{R}}^*,S_P,S_C) = \mathcal{R}_r(\bm{W_{L}},\bm{W_{R}},S_L,S_R,Q_{lat},\sigma,\kappa,j),
    \label{Eq:reduced_system}
\end{equation}
where $j$ is the only unknown variable on the right side. In this paper, the adjacent states of the phase interface $\bm{W_{liq}}^*$ and $\bm{W_{vap}}^*$ are used to calculate the mass flux (see Eq. \eqref{eq:mass_flux}), while the initial states $\bm{W_{L}}$ and $\bm{W_{R}}$ are used in Refs. \cite{houim2013ghost,das2020sharp}. In fact, as shown in Eq. \eqref{Eq:reduced_system}, we can obtain the approximate Riemann solution non-iteratively if $j$ is evaluated with $\bm{W_{L}}$ and $\bm{W_{R}}$. However, as we will show in Section. 4.1, this simple strategy will lead to an inconsistent numerical solution, which is not discussed in the previous researches. With the relations $\bm{W_{liq}}^* = \bm{W_{liq}}^*(j)$ and $\bm{W_{vap}}^* = \bm{W_{vap}}^*(j)$, the target equation Eq. \eqref{Eq:phase_change_model} is rewritten as $j = f_m(j)$ so that it can be solved with a standard root-finding algorithm. In this paper, we employ the steffense method \cite{johnson1968steffensen}, which is formulated as 
\begin{equation}
    \begin{aligned}
    j_{k+1} &= \psi(j_k),\quad (k = 0, 1, 2,...),\\
    \psi(j) &= j - \frac{[f_m(j) - j]^2}{f_m(f_m(j)) - 2f_m(j) + j},
    \end{aligned}    
\end{equation}
where the initial value $j_0$ is estimated by using $\bm{W_{L}}$ and $\bm{W_{R}}$. This iteration procedure will be terminated when $|j_{k+1} - j_{k}| <10^{-6}$ and 3-4 iterations are typically enough. Nevertheless, since the Euler equation is used and the heat conduction is neglected, the temperature jump at the contact wave can be arbitrary large. As reported in Ref. \cite{fechter2017sharp}, being the potential source of numerical instability, unphysical large temperature jumps will be predicted at the contact wave in some complex problems. To ensure numerical stability, we add an additional constraint for the temperature difference at the contact wave. When it is larger than $50\ \rm{K}$, the iteration will be aborted with the mass flux fixed at $j = j_0$. 

After obtaining the approximate Riemann solution, the interface velocity is set as $S_p$. Following Ref. \cite{lin2017simulation}, we extend the interface velocity to the cells in the vicinity of the interface along the normal direction. In addition, the interfacial fluxes can be computed by
\begin{equation}
    \hat{\bm{X}}_{vap/liq}= \pm \Delta \Gamma \left[
    \begin{aligned}
    &j \\
    (&j\ V_{vap/liq}^* + p_{vap/liq}^*)n_x \\ 
    (&j\ V_{vap/liq}^* + p_{vap/liq}^*)n_y \\
    (&j\ V_{vap/liq}^* + p_{vap/liq}^*)n_z \\
    &j\ (e_{vap/liq}^* + \frac{1}{2} {V_{vap/liq}^{*2}} ) + p_{vap/liq}^*V_{vap/liq}^*  \\
    \end{aligned}
    \right],
    \label{Eq:interfacial_fluxes}
\end{equation} 
where $+$ and $-$ are applied for the vapor phase and the liquid phase, respectively. The conservation property of the present method is confirmed by the relation as follows,
\begin{equation}
    \hat{\bm{X}}_{vap} + \hat{\bm{X}}_{liq} = [0,-\sigma\kappa n_x,-\sigma\kappa n_y,-\sigma\kappa n_z,jQ_{lat} - \sigma\kappa S_p ]^T.
\end{equation}
Note that the momentum conservation and the energy conservation are not satisfied, and the reason is twofold. First, the pressure jump caused by surface tension leads to the violation of the momentum conservation and the energy conservation. However, as the present method is based on the conservative sharp-interface
framework, it maintains the so called zero-order consistency \cite{luo2015conservative}, i.e., the total force acting on the drop with constant curvature is zero. Second, to model the phase interface in a thermodynamic consistent way with the heat conduction neglected \cite{fechter2018approximate}, the latent heat is introduced in the energy jump condition as a source term. The phase interface will liberate thermal energy in the evaporation process while the opposite situation happens in the condensation process, which are known as the heat of evaporation and condensation, respectively. When phase change and surface tension are not considered, the interfacial fluxes Eq. \eqref{Eq:interfacial_fluxes} will recover the original scheme of Hu et al. \cite{hu2006conservative}.

\section{Numerical examples}
\label{sec4}
In this section, various numerical examples are presented to validate the present method. The inviscid term is discretized by a fifth-order accurate WENO scheme \cite{shu1988efficient}, which is also used to solve the level-set advection equation. The time marching of both the fluid equation and level-set equation is performed by the second-order accurate SSP Runge-Kutta
scheme [36]. For all cases, the Courant-Friedrichs-Lewy (CFL) number is set to 0.6. Moreover, a wavelet-based adaptive
multi-resolution (MR) algorithm \cite{han2014adaptive} is employed to improve computational efficiency.

\subsection{One-dimensional validation}

\subsubsection{Riemann problem of n-dodecane with evaporation}
\label{subsubsec4.1.1}
\begin{figure}[htbp]
    \centering
    \includegraphics[width=1.0\textwidth]{ 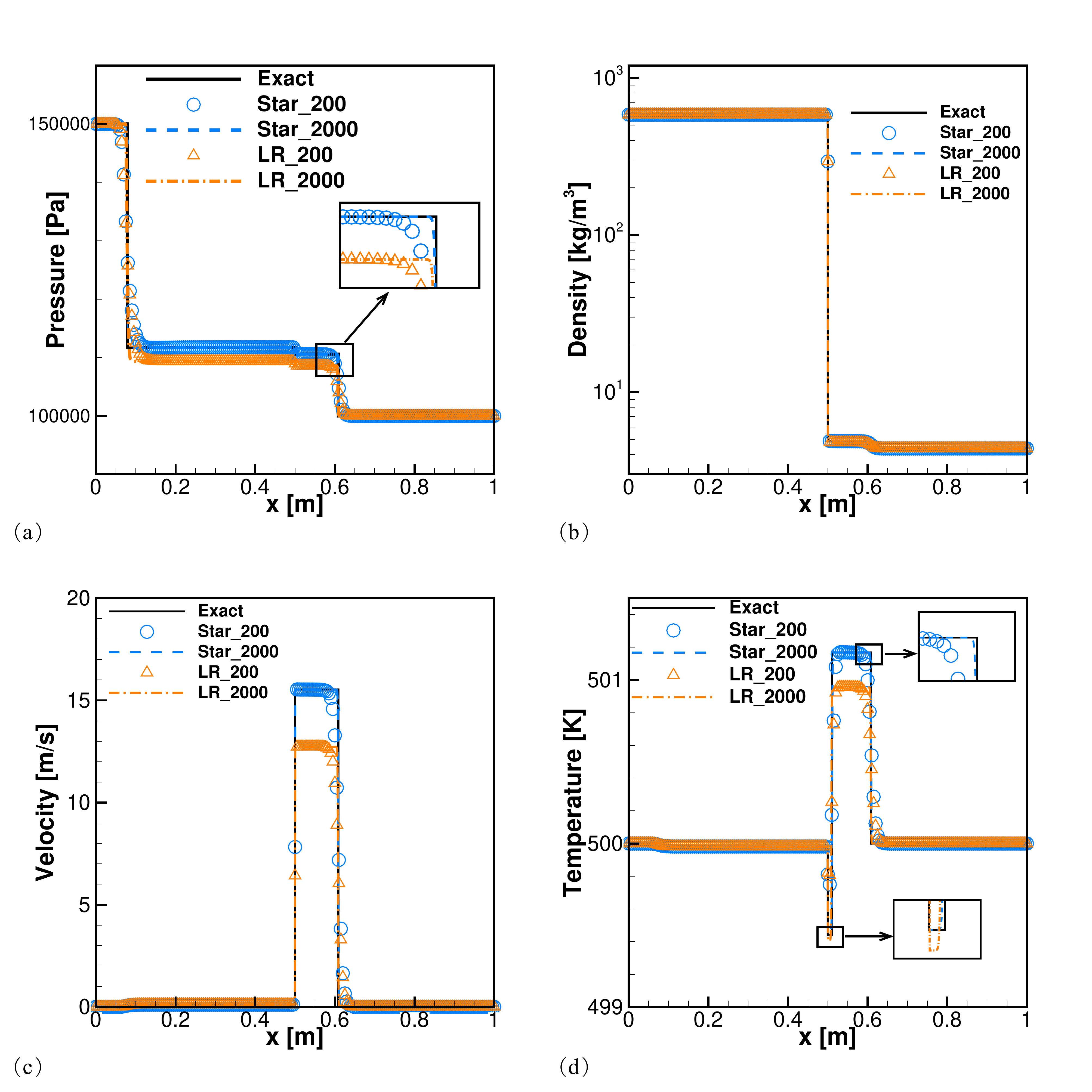}
    \caption{Riemann problem of n-dodecane with evaporation:  (a) the pressure profile, (b) the density profile, (c) the velocity profile, and (d) the temperature file. The simulations are carried out with the mass flux calculated by $j = f_m(\bm{W_{liq}}^*,\bm{W_{vap}}^*)$ and $j = f_m(\bm{W_{L}},\bm{W_{R}})$, which are denoted by ``Star'' and ``LR'', respectively. The results are obtained on the uniform grids involving $200$ and $2000$ cells.}
    \label{Fig:dodecane_evap}
\end{figure}
Taken from \cite{fechter2017sharp,fechter2018approximate}, the Riemann problem of n-dodecane with resolved evaporation effects is considered to validate the present method. The initial conditions are given as
\begin{equation}
(\rho,u,p,T)=\left\{
\begin{aligned}
&(584.08\ \rm{kg/m^3},0\ \rm{m/s},1.5\times10^5\ \rm{Pa},500\ \rm{K})\  &\text{if $x\leq0.5$ m,}\\
&(4.38\ \rm{kg/m^3},0\ \rm{m/s},1.0\times10^5\ \rm{Pa},500\ \rm{K})\  &\text{if $x>0.5$ m,}\\
\end{aligned}
\right.
\label{Eq:dodecane_evap_initial_condition}
\end{equation}
and the final time is $t = 0.7\ \rm{ms}$. As mentioned above, the mass flux can be evaluated through the adjacent states of the phase interface ($\bm{W_{liq}}^*$ and $\bm{W_{vap}}^*$) or the initial states ($\bm{W_{L}}$ and $\bm{W_{R}}$). The results obtained with the two strategies are given in Fig. \ref{Fig:dodecane_evap}, which are labeled by ``Star'' and ``LR'', respectively. Note that there is a common exact solution for the two situations, which is achieved via setting the model coefficients in Eq. \eqref{Eq:HKR} as 
\begin{equation*}
\left\{
\begin{aligned}
&(\lambda_{evap}, \lambda_{cond})^{Star} = (1.0,0.9),\\ 
&(\lambda_{evap}, \lambda_{cond})^{LR} = (0.756,0.680).
\end{aligned}
\right.
\end{equation*}
The uniform grids with $200$ and $2000$ cells are used to investigate the convergence of the present method. It can be observed from Fig. \ref{Fig:dodecane_evap} that, by employing $j = f_m(\bm{W_{liq}}^*,\bm{W_{vap}}^*)$, the numerical results agree well with the exact solution. In contrast, the results obtained with $j = f_m(\bm{W_{L}},\bm{W_{R}})$ will converge to an incorrect solution. This issue is not even noticed in the previous researches, and it will also occur in the condensation case. For simplicity, the evaporation case is chosen to elucidate this issue.

Let $\mathcal{S}$ denotes the unknown states in the star region. Then the exact Riemann solution can be expressed as 
\begin{equation*}
\mathcal{S}(\bm{W_{L}}^{*e},\bm{W_{M}}^{*e},\bm{W_{R}}^{*e}) = \mathcal{R}(\bm{W_{L}}^0,\bm{W_{R}}^0),
\end{equation*}
where the superscript $0$ represents the initial conditions given in Eq. \eqref{Eq:dodecane_evap_initial_condition}. Therefore, in the numerical simulation, the initial states of the interfacial Riemann problem are $\bm{W_{L}}^0$ and $\bm{W_{R}}^0$ at the beginning and then change to $\bm{W_{L}}^{*e}$ and $\bm{W_{M}}^{*e}$ within a few time steps. If they are used to evaluate the mass flux, an inconsistent numerical solution will arise due to $f_m(\bm{W_{L}}^0,\bm{W_{R}}^0) \neq f_m(\bm{W_{L}}^{*e},\bm{W_{M}}^{*e})$. Conversely, the mass flux obtained by using the adjacent states of the phase interface is consistent during the whole simulation since the relation
\begin{equation}
    \mathcal{S}(\bm{W_{L}}^{*e},\bm{W_{M}}^{*e},\bm{W_{M}}^{*e}) = \mathcal{R}(\bm{W_{L}}^e,\bm{W_{M}}^{*e})
\end{equation} 
holds. In conclusion, no matter which phase change model is adopted, the mass flux should be calculated through the adjacent states of the phase interface to ensure numerical consistency.

\subsubsection{Riemann problem of n-dodecane with condensation}
\label{subsubsec4.1.2}
\begin{figure}[htbp]
    \centering
    \includegraphics[width=1.0\textwidth]{ 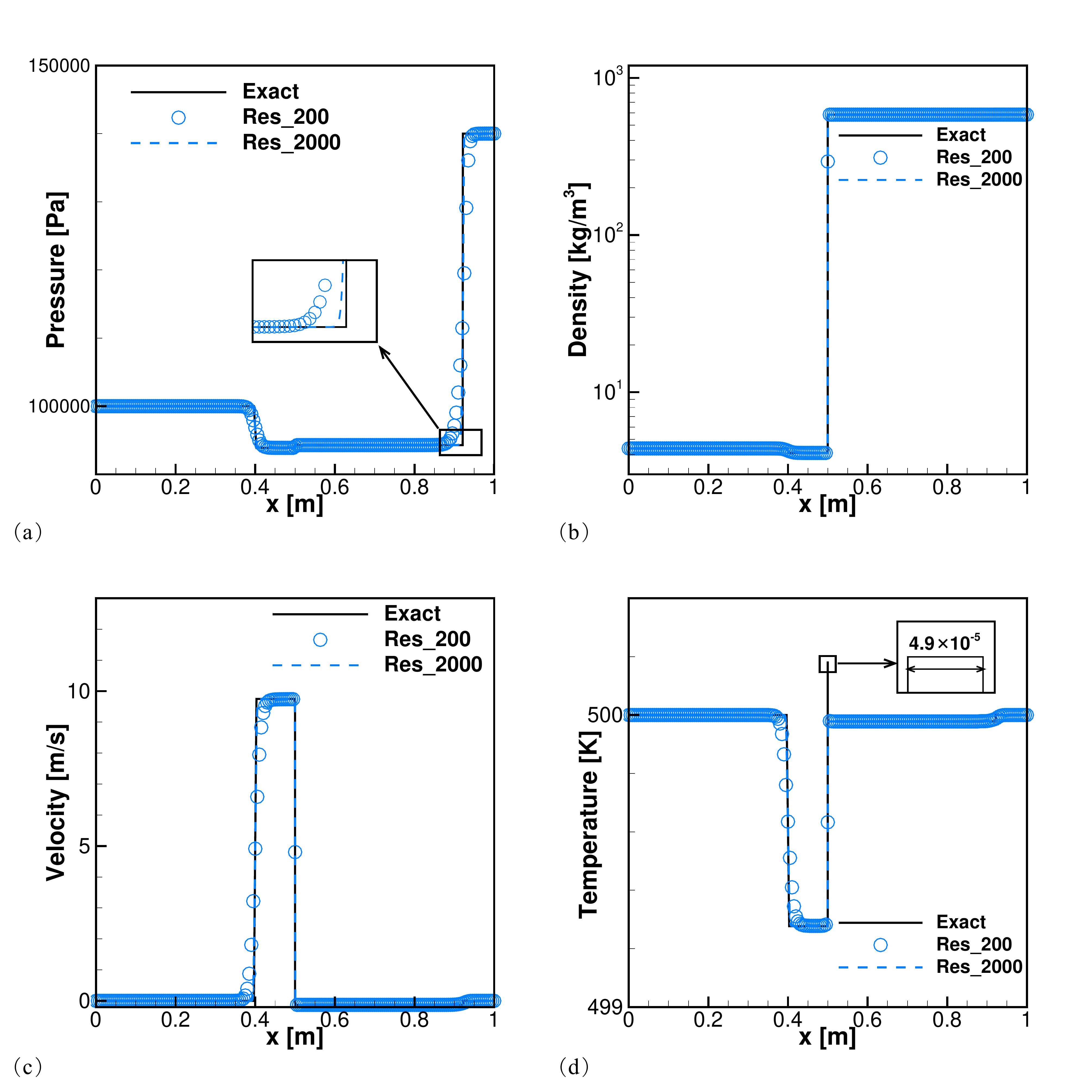}
    \caption{Riemann problem of n-dodecane with condensation:  (a) the pressure profile, (b) the density profile, (c) the velocity profile, and (d) the temperature file. The simulations are carried out on the uniform grids involving $200$ and $2000$ cells.}
    \label{Fig:dodecane_cond}
\end{figure}
We consider the Riemann problem of n-dodecane with condensation here. With $\lambda_{evap} = 0.6$ and $\lambda_{cond} = 1.0$, the initial conditions are given by
\begin{equation}
(\rho,u,p,T)=\left\{
\begin{aligned}
&(4.38\ \rm{kg/m^3},0\ \rm{m/s},1.0\times10^5\ \rm{Pa},500\ \rm{K})\  &\text{if $x\leq0.5$ m,}\\
&(584.05\ \rm{kg/m^3},0\ \rm{m/s},1.4\times10^5\ \rm{Pa},500\ \rm{K})\  &\text{if $x>0.5$ m,}\\
\end{aligned}
\right.
\label{Eq:dodecane_cond_initial_condition}
\end{equation}
and the final time is $t = 0.7\ \rm{ms}$. As shown in Fig. \ref{Fig:dodecane_cond}, the outer waves are two rarefaction waves, near which the smearing due to numerical diffusion is observed. Similarly, the computational domain is discretized by the uniform grids involving $200$ and $2000$ cells. With a higher resolution, the more sharp rarefaction waves are obtained. However, unlike in the evaporation case, the interval between the phase interface and the contact wave is not resolved in the numerical results. This is because its length scale is much smaller when condensation is considered. In the present problem, the length of this interval is $4.9\times10^{-5} \ \rm{m}$, and about $20000$ grid cells are needed to resolve it.

\subsubsection{Riemann problem of water with evaporation}
\label{subsubsec4.1.3}
\begin{figure}[htbp]
    \centering
    \includegraphics[width=1.0\textwidth]{ 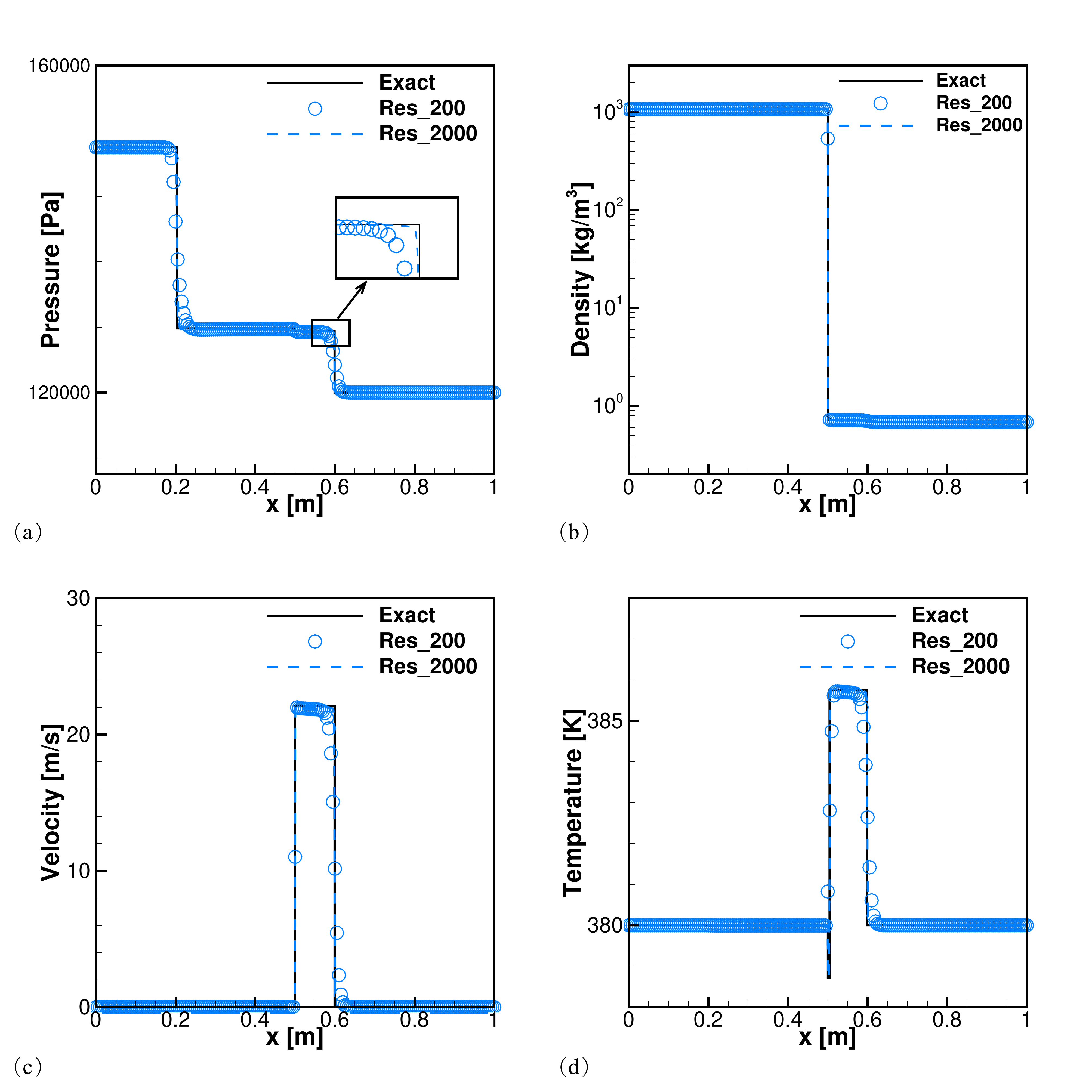}
    \caption{Riemann problem of water with evaporation:  (a) the pressure profile, (b) the density profile, (c) the velocity profile, and (d) the temperature file. The simulations are carried out on the uniform grids involving $200$ and $2000$ cells.}
    \label{Fig:water_evap}
\end{figure}

The Riemann problem with evaporation is simulated for water governed by the stiffened gas EOS. The initial conditions are
\begin{equation}
    (\rho,u,p,T)=\left\{
    \begin{aligned}
         & (1073.57\ \rm{kg/m^3},0\ \rm{m/s},1.5\times10^5\ \rm{Pa},380\ \rm{K})\    & \text{if $x\leq0.5$ m,} \\
         & (0.68\ \rm{kg/m^3},0\ \rm{m/s},1.2\times10^5\ \rm{Pa},380\ \rm{K})\  & \text{if $x>0.5$ m,}    \\
    \end{aligned}
    \right.
    \label{Eq:water_evap_initial_condition}
\end{equation}
and the final time is $t = 0.3\ \rm{ms}$. With $\lambda_{evap} = 1.0$ and $\lambda_{cond} = 0.9$, the simulations are carried out on the uniform grids involving $200$ cells and $2000$ cells, see Fig. \ref{Fig:water_evap}. As in the evaporation of n-dodecane, from left to right, the solution consists of a rarefaction wave, an evaporation wave, a contact wave and a shock wave. The interval between the evaporation wave and the contact wave is not resolved by using $200$ grid points due to its small length scale introduced by the stiffness of water. With $2000$ grid cells, all of the waves are correctly reproduced in the numerical results.

\subsubsection{Riemann problem of water with condensation}
\label{subsubsec4.1.4}
\begin{figure}[htbp]
    \centering
    \includegraphics[width=1.0\textwidth]{ 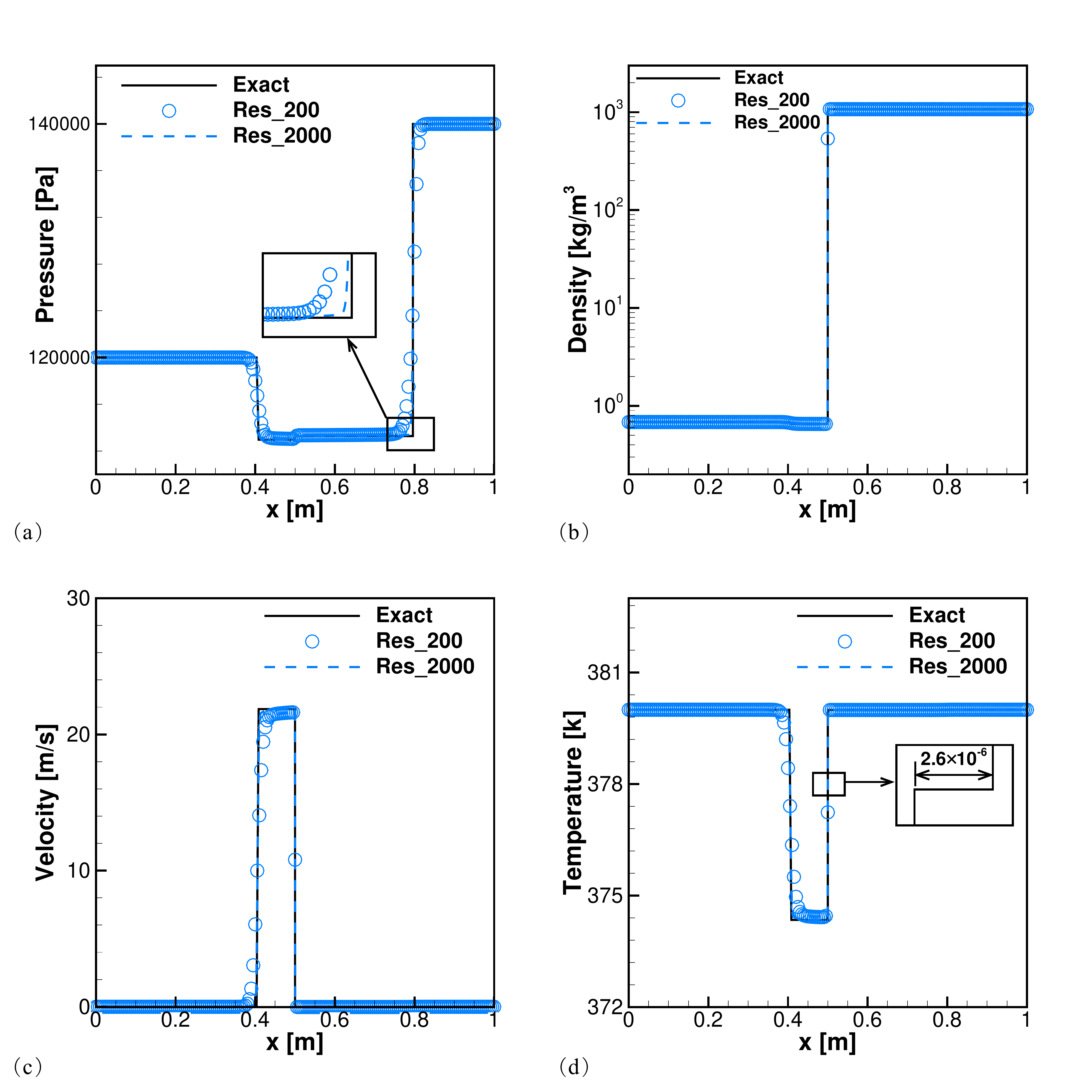}
    \caption{Riemann problem of water with condensation:  (a) the pressure profile, (b) the density profile, (c) the velocity profile, and (d) the temperature file. The simulations are carried out on the uniform grids involving $200$ and $2000$ cells.}
    \label{Fig:water_cond}
\end{figure}
Here we consider the Riemann problem of water with condensation. The initial conditions are given by
\begin{equation}
    (\rho,u,p,T)=\left\{
    \begin{aligned}
         & (0.68\ \rm{kg/m^3},0\ \rm{m/s},1.2\times10^5\ \rm{Pa},380\ \rm{K})\    & \text{if $x\leq0.5$ m,} \\
         & (1073.56\ \rm{kg/m^3},0\ \rm{m/s},1.4\times10^5\ \rm{Pa},380\ \rm{K})\  & \text{if $x>0.5$ m,}    \\
    \end{aligned}
    \right.
    \label{Eq:water_cond_initial_condition}
\end{equation}
and the final time is $t = 0.3\ \rm{ms}$. By setting the model coefficients as $\lambda_{evap} = 0.8$ and $\lambda_{cond} = 1.0$, the water vapor will condense steadily. The computational domain is discretized by the uniform grids containing $200$ and $2000$ cells. As shown in Fig. \ref{Fig:water_cond}, the numerical results are in good agreement with the exact solution. Being far smaller than the grid cell size, the length of the interval between the contact wave and the condensation wave is only $2.6 \times 10^{-6}\ \rm{m}$ so that it is not resolved in the numerical results.

\subsection{Two-dimensional validation}
\subsubsection{Circular droplet with evaporation}
\label{subsubsec4.2.1}
\begin{figure}[htbp]
    \centering
    \includegraphics[width=1.0\textwidth]{ 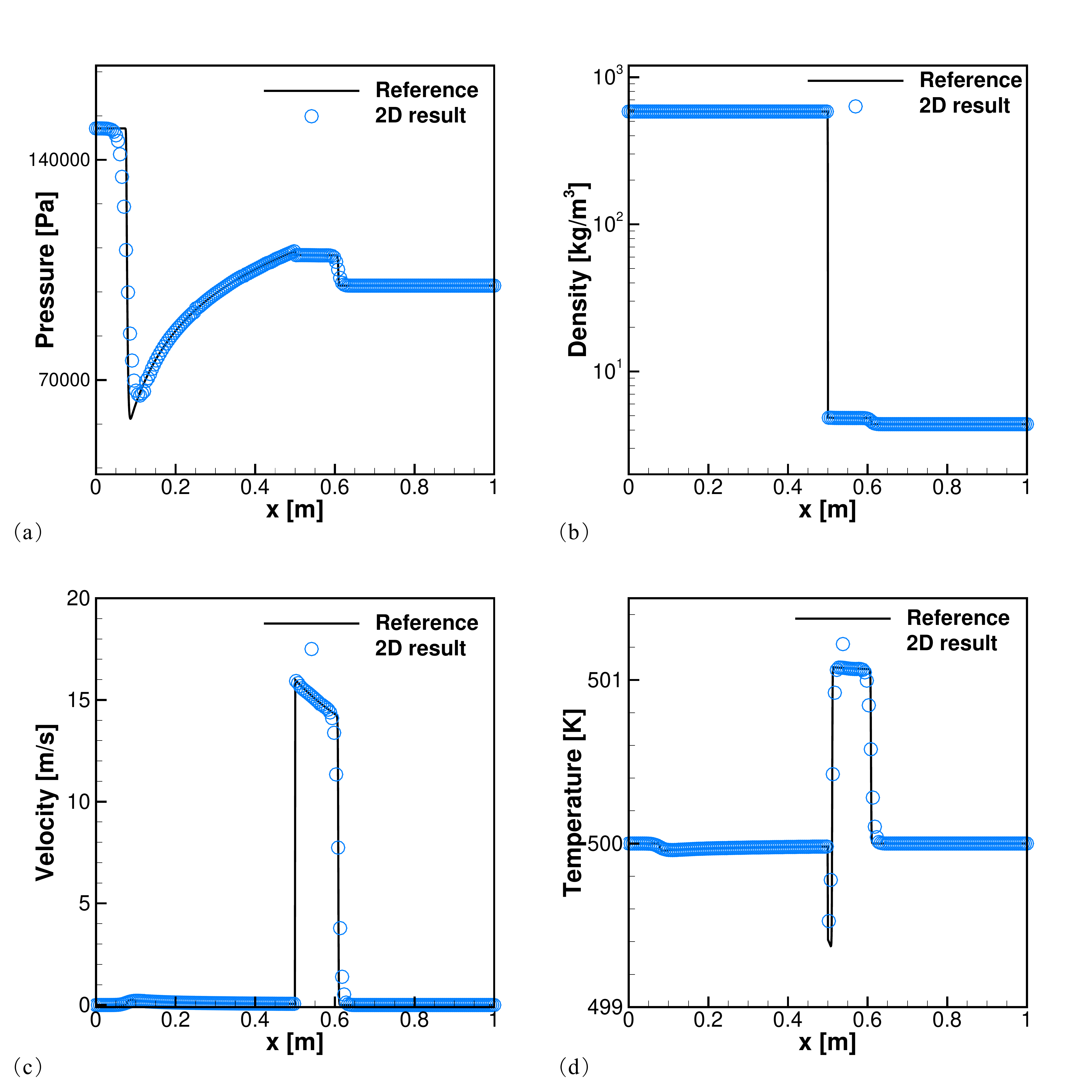}
    \caption{Circular droplet with evaporation:  (a) the pressure profile, (b) the density profile, (c) the velocity profile, and (d) the temperature file. The 2D simulation is carried out on a MR grid with an effective resolution of $256\times256$ and the results shown here are extracted along the x-axis. With $2000$ grid points, the reference solution is obtained by employing the radially symmetric one-dimensional approach \cite{toro2013riemann}.}
    \label{Fig:dodecane_evap_2D}
\end{figure}
This case is a two-dimensional extension of the one-dimensional Riemann problem of n-dodecane with evaporation. The initial states of the vapor phase and the liquid phase are the same as in Section. \ref{subsubsec4.1.1}, and the level-set is initialized by
\begin{equation}                   
    \phi  =-0.5+\sqrt{x^{2}+y^{2}}\quad \rm{m}. 
\label{Eq:2d_evap_levelset}
\end{equation}
The computational domain is a unit square with symmetry boundary conditions. Note that only $1/4$ of the droplet is considered here, whose center coincides with the origin. With $\lambda_{evap} = 1.0$ and $\lambda_{cond} = 0.9$, the two-dimensional simulation is carried on a MR grid where the effective resolution at the finest level is $256\times256$. By employing a radially symmetric approach \cite{toro2013riemann}, a one-dimensional simulation is conducted with $2000$ grid points and is regarded as reference solution. The final time is $t = 0.7\ \rm{ms}$, and the results are shown in Fig. \ref{Fig:dodecane_evap_2D}. It can be observed that, influenced by the geometry, the states between those waves are no longer constant. Even with a much coarser resolution, the pressure, density, velocity, and temperature profiles of the two-dimensional simulation, which are extracted along the x-axis, fit the reference solution very well.

During the simulation, the mass of each phase and the total mass can be calculated by $\bm{M}_{vap} = \sum_i \alpha_i \rho_i^{vap}$, $\bm{M}_{liq} = \sum_i (1 - \alpha_i)\rho_i^{liq}$, and $\bm{M}_{total} = \bm{M}_{vap} + \bm{M}_{liq} $, respectively.
To measure the mass variations, the relative mass is defined as
\begin{equation}
    \overline{\bm{M}}^{n}_{liq/vap/total} = \frac{\bm{M}^n_{liq/vap/total}}{\bm{M}^0_{liq/vap/total}},
\end{equation}
where the superscripts $0$ and $n$ represent the initial condition and the states at the $n$-th time step. Fig. \ref{Fig:conservation_2D}(a) plots the relative mass of each phase versus simulation time, in which the mass transfer from the liquid phase to the vapor phase can be observed. The conservation of the present method is confirmed by the total relative mass equaling to one during the whole simulation, see Fig. \ref{Fig:conservation_2D}(a).

\subsubsection{Circular bubble  with condensation}
\label{subsubsec4.2.2}
\begin{figure}[htbp]
    \centering
    \includegraphics[width=1.0\textwidth]{ 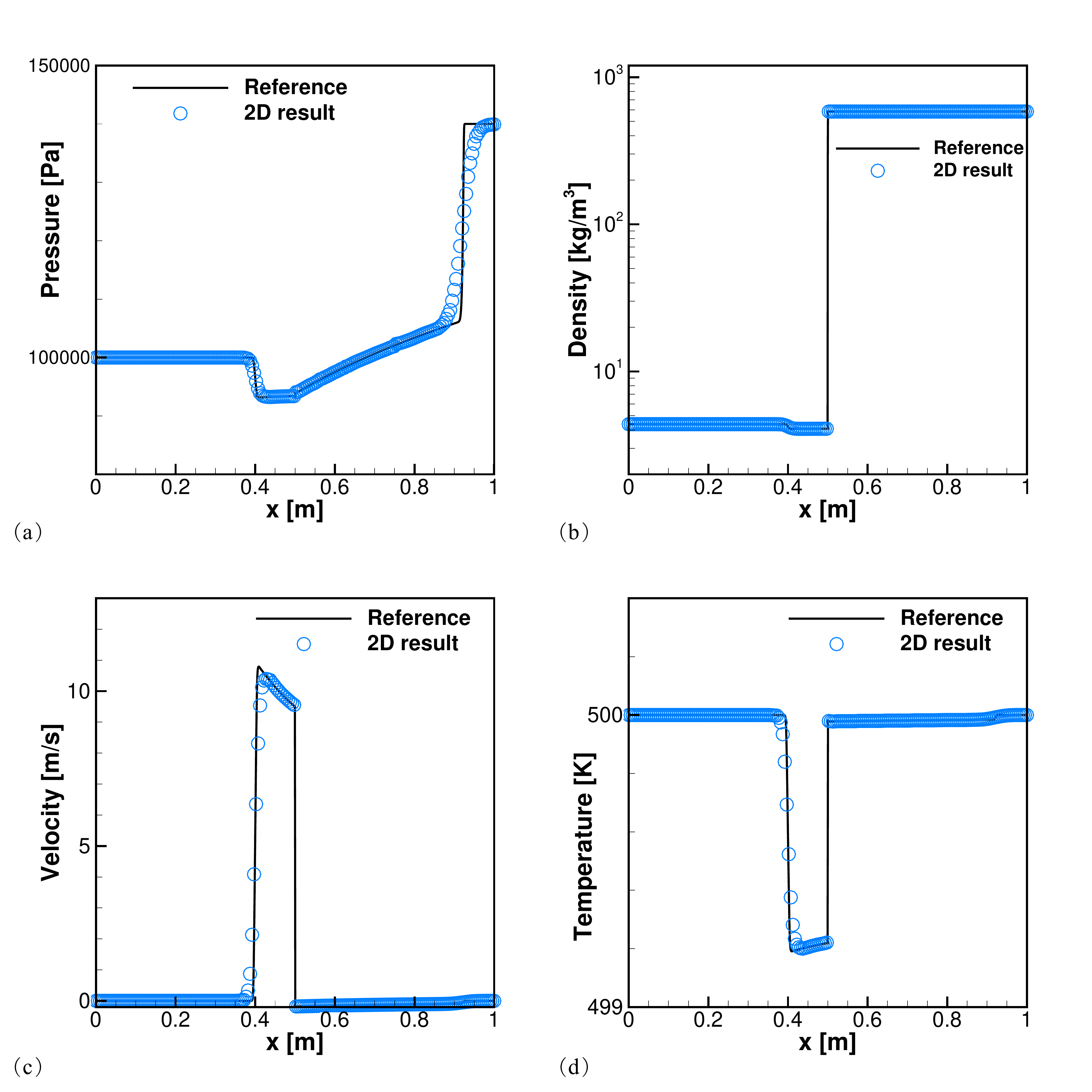}
    \caption{Circular bubble with condensation:  (a) the pressure profile, (b) the density profile, (c) the velocity profile, and (d) the temperature file. The 2D simulation is carried out on a MR grid with an effective resolution of $256\times256$ and the results shown here are extracted along the x-axis. With $2000$ grid points, the reference solution is obtained by employing the radially symmetric one-dimensional approach \cite{toro2013riemann}.}
    \label{Fig:dodecane_cond_2D}
\end{figure}
\begin{figure}[htbp]
    \centering
    \includegraphics[width=1.0\textwidth]{ 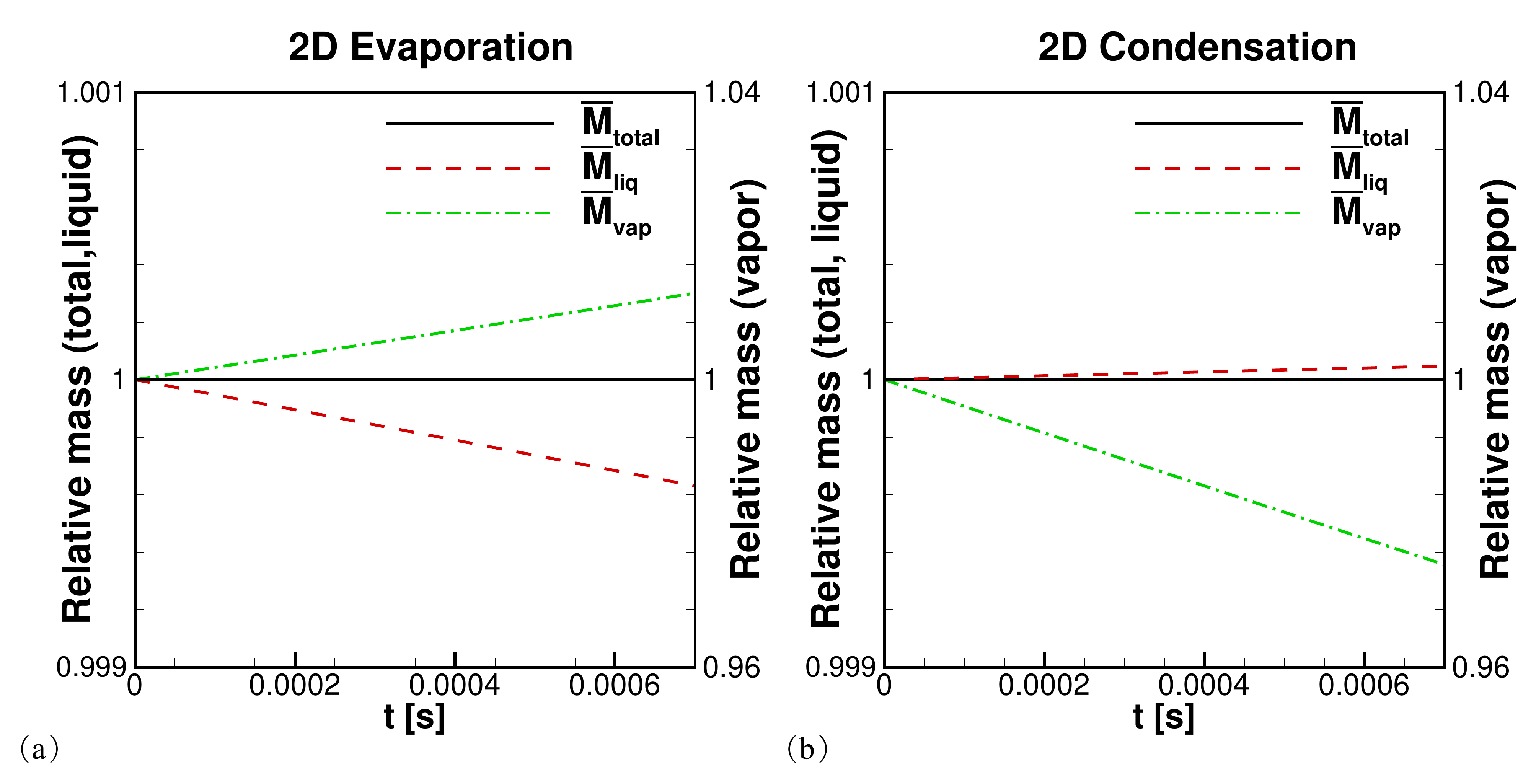}
    \caption{Two-dimensional validation: the mass variations versus the simulation time.}
    \label{Fig:conservation_2D}
\end{figure}

The two-dimensional condensation of an n-dodecane bubble is computed here. As in Section. \ref{subsubsec4.2.1}, we consider only one-fourth of a circular bubble with a radius $R = 0.5\ \rm{m}$. It is placed in the lower left corner of a $1\ \rm{m} \times 1\ \rm{m}$ square computational domain with symmetry boundary conditions. The initial states, the model coefficients, and the final time are the same as in Section. \ref{subsubsec4.1.2}. Similarly, the cut along the x-axis of the two-dimensional results is compared with one-dimensional reference solution obtained on $2000$ grid points, see Fig. \ref{Fig:dodecane_cond_2D}. The multi-dimensional implementation of the present method reproduces well the results of the one-dimensional approach assuming radially symmetry \cite{toro2013riemann}. Note that the two-dimensional results are obtained with a coarse resolution of only $256\times256$. The mass transfer from the vapor phase to the liquid phase can be observed from Fig. \ref{Fig:conservation_2D}(b), which also confirms that the present method has no conservation error.

\subsection{Oscillating droplet}
\begin{figure}[htbp]
    \centering
    \includegraphics[width=1.0\textwidth]{ 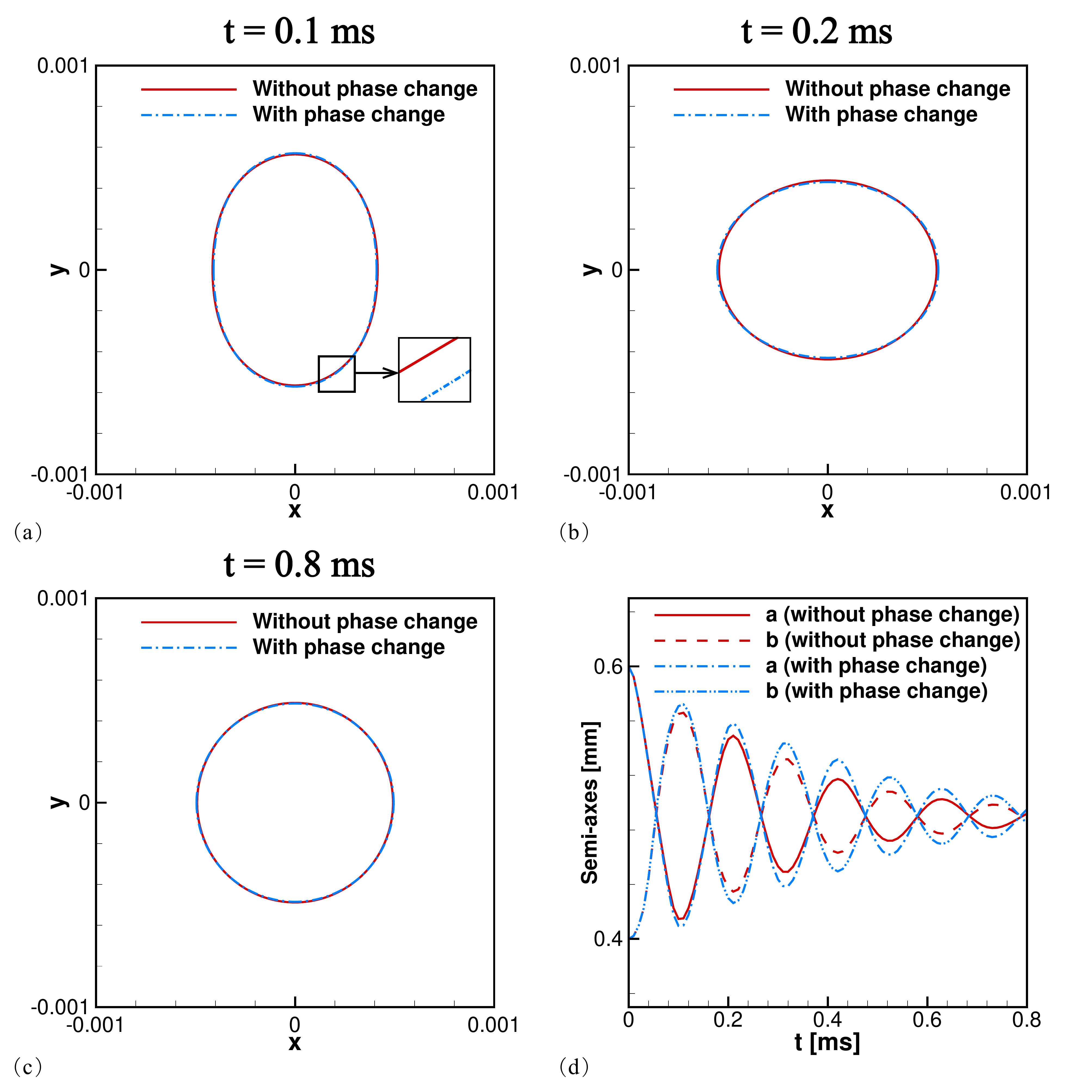}
    \caption{Oscillating droplet: (a)-(c) the shapes of the droplet at different time instants, and (d) the evolution of the elliptical semi-axes. To utilize the symmetry of this problem, only $1/4$ of the physical domain is computed with an effective resolution of $256 \times 256$.}
    \label{Fig:oscillating_drop}
\end{figure}
As in Refs. \cite{fechter2017sharp,fechter2018approximate}, the oscillating droplet case is considered to validate the present method for problems dominated by surface tension. Initially, with the semi-axes $a = 0.6\ \rm{mm}$ and $b = 0.4\ \rm{mm}$, an elliptical droplet is placed at the center of a $5\ \rm{mm} \times 5\ \rm{mm}$ square domain with zero kinetic energy. Due to surface tension, this droplet will oscillate periodically along with the translation between the potential energy and the kinetic energy. The symmetry property of the present case is utilized by computing only $1/4$ of the physical domain. We set the surface tension coefficient as $\sigma = 20\ \rm{Nm^{-1}}$, and the initial conditions as
\begin{equation}
    \left\{
    \begin{aligned}
        \rho & =0.74\ \rm{kg/m^{3}}, u = 0\ \rm{m/s}, v = 0\ \rm{m/s} \quad \text{vapor} \\
        p    & =1.3\times10^{5}\ \rm{Pa}, T=380\ \rm{K}                                       \\
        \rho & =1073.55\ \rm{kg/m^{3}}, u = 0\ \rm{m/s}, v = 0\ \rm{m/s} \quad \text{droplet} \\
        p    & =1.3\times10^{5}\ \rm{Pa}, T=380\ \rm{K}                                       \\
        \phi & =-1.0+\sqrt{(x/0.6)^{2}+(y/0.4)^{2}}\ \rm{mm}\quad level\ set.
    \end{aligned}
    \right.
    \label{Eq:oscillating_iniital_condition}
\end{equation}
Two simulations with and without phase change are conducted on a MR grid with an effective resolution of $256\times256$. The model coefficients are set as $\lambda_{evap} = \lambda_{cond} = 1.0$. Note that the approximate Riemann solver is performed with $j = 0$ when phase change is not considered, see Appendix. A for details. Fig. \ref{Fig:oscillating_drop}(a)-(c) depict the shapes of the droplet at different time instants. Qualitatively, no obvious differences are observed between the interfaces obtained with and phase change, as the phase change effects are small according to the initial conditions. For a quantitative analysis, the lengths of the semi-axes are plotted over time in Fig. \ref{Fig:oscillating_drop}(d). With the phase change effects, the oscillation amplitude is changed by the additional forces acting on the surface while the oscillation frequency remains the same. According to the Rayleigh formula extended to two-phase flows \cite{fyfe1988surface,long2021accelerated}, the analytical oscillation period can be calculated by
\begin{equation}
    \omega^2 = (o^3 - o)\frac{\sigma}{(\rho_{liq} +\rho_{vap})R^3},\quad T = \frac{2 \pi}{\omega},
    \label{Eq:oscillation_period}
\end{equation}
where $o$ is the oscillation mode and $R$ is the radius of the liquid drop at the equilibrium state. In this case, we have $o = 2$ and $R = 0.4899\ \rm{mm}$ so that the theoretical oscillation period is $T = 0.204\ \rm{ms}$. As shown in Fig. \ref{Fig:oscillating_drop}(d), the oscillation period obtained by the present method is $T = 0.0881$ s, i.e. an error of about $0.3\%$.

\subsection{Shock-Droplet interaction}
\subsubsection{Two-dimensional shock and n-dodecane droplet interaction}
\label{subsubsec4.4.1}
\begin{figure}[htbp]
    \centering
    \includegraphics[width=1.0\textwidth]{ 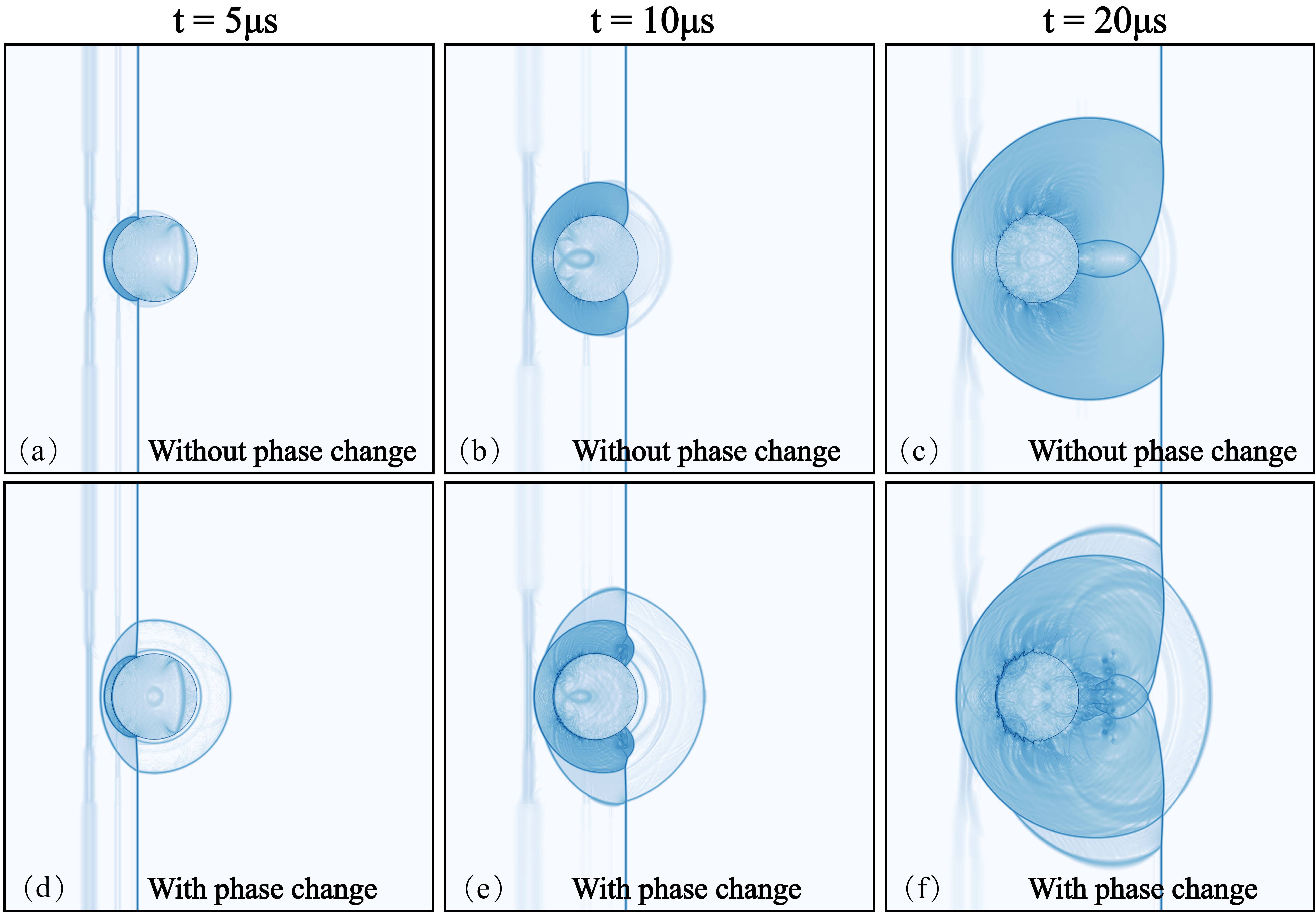}
    \caption{Two-dimensional shock and n-dodecane droplet interaction: the numerical Schlieren images calculated by $\log{(|\nabla \rho| +1)}$. Each column shows the results at the same time instant with and without phase change. Assuming that the flow is symmetric about x-axis, only the top half domain is computed with an effective resolution of $2048\times1024$.}
    \label{Fig:shock_drop_dodecane_1}
\end{figure}
\begin{figure}[htbp]
    \centering
    \includegraphics[width=1.0\textwidth]{ 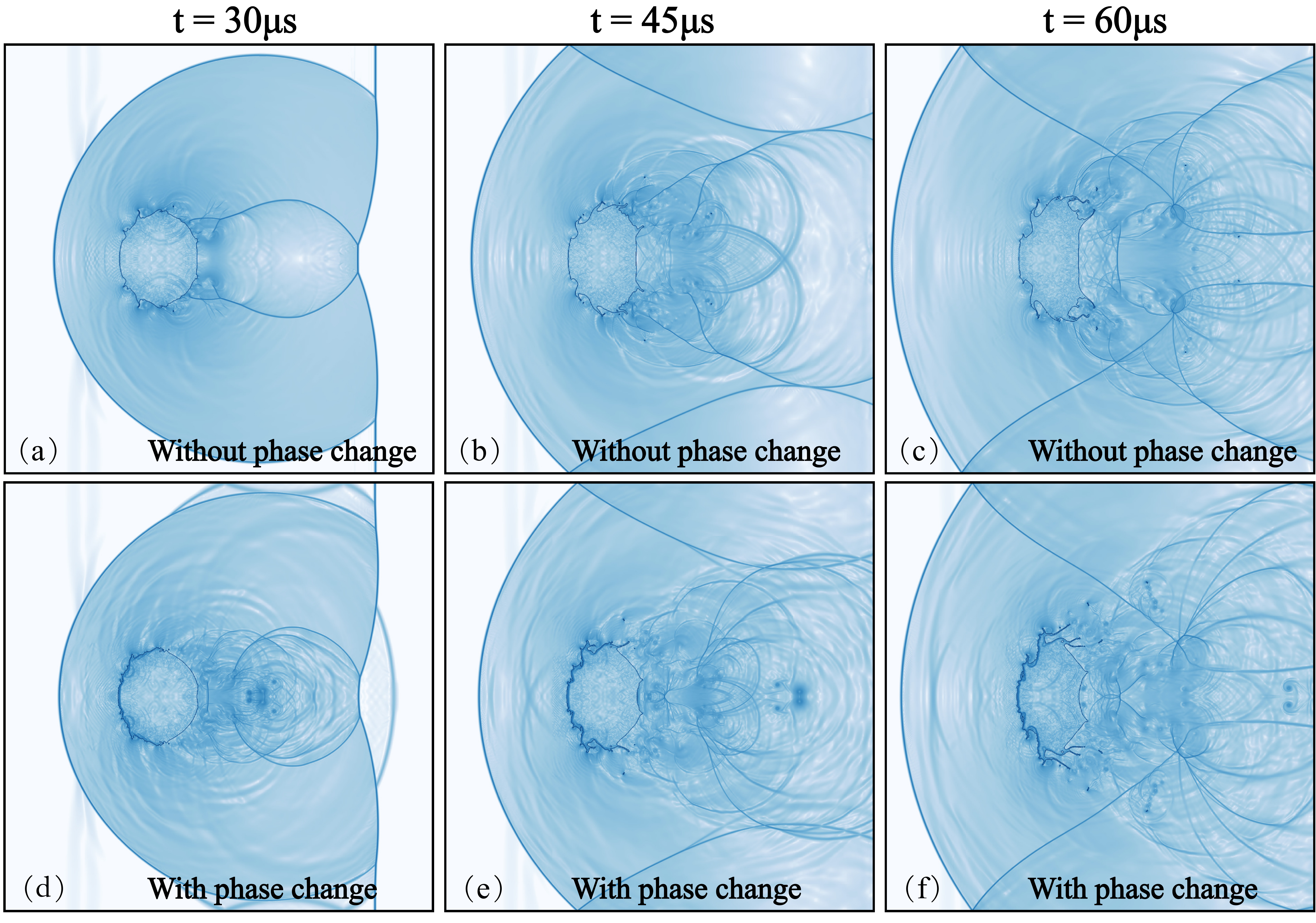}
    \caption{Two-dimensional shock and n-dodecane droplet interaction: the numerical Schlieren images calculated by $\log{(|\nabla \rho| +1)}$. Each column shows the results at the same time instant with and without phase change. Assuming that the flow is symmetric about x-axis, only the top half domain is computed with an effective resolution of $2048\times1024$.}
    \label{Fig:shock_drop_dodecane_2}
\end{figure}
Following Refs. \cite{fechter2017sharp,fechter2018approximate}, we consider the problem that a planar shock wave impinging upon a cylindrical n-dodecane droplet. The original computational domain is a square of $10\ \rm{mm} \times 10\ \rm{mm}$ with a shock wave of Mach number $1.5$ placed at $x = 2\ \rm{mm}$. To improve efficiency, the flow filed is assumed to be symmetric in the y-direction and only the top half domain is computed with an effective resolution of $2048\times1024$. The initial conditions are given by 
\begin{equation}
    \left\{
    \begin{aligned}
        \rho & =4.38\ \rm{kg/m^3}, u=0\ \rm{m/s},v=0\ \rm{m/s},   \quad \text{pre-shocked vapor},  \\
        p&=1.0\times10^{5}\ \rm{Pa}, T=500\ \rm{K}\\
        \rho & =10.63\ \rm{kg/m^3}, u=130.29\ \rm{m/s},v=0\ \rm{m/s},   \quad \text{post-shocked vapor},  \\
        p&=226613\ \rm{Pa}, T=511.19\ \rm{K}\\
        \rho & =593.47\ \rm{kg/m^3}, u=0\ \rm{m/s},v=0\ \rm{m/s},   \quad \text{n-dodecane droplet},  \\
        p&=1.0\times10^{5}\ \rm{Pa}, T=490\ \rm{K}\\
        \phi & =-1.0+\sqrt{(x-0.35)^{2}+y^{2}} \ \rm{mm}        \quad level\ set.
    \end{aligned}
    \right.
\end{equation}
Symmetry boundary conditions are employed at the top and the bottom sides, while the left and the right sides are set as inflow and outflow boundaries, respectively. With $\lambda_{evap} = 1.0$ and $\lambda_{cond} = 0.6$, the droplet will evaporate in the pre-shock state. Serving as a contrast, the problem without phase change is also computed via fixing the mass flux at zero. The results at different time instants are given in Figs. \ref{Fig:shock_drop_dodecane_1} and \ref{Fig:shock_drop_dodecane_2}. As in Refs. \cite{fechter2017sharp,fechter2018approximate}, an additional shock wave is captured when phase change is considered, which is caused by the initial evaporation of the droplet. After the shock wave hits the droplet, condensation is also detected on the droplet surface as the surrounding vapor conditions are changed. In compare to the situation without phase change, more complex flow structures and larger interface deformations are observed in the results with resolved phase change effects. Note that our results show much more details than the previous simulations \cite{fechter2017sharp,fechter2018approximate}. 

\subsubsection{Two-dimensional shock and water droplet interaction}
\label{subsubsec4.4.2}
\begin{figure}[htbp]
    \centering
    \includegraphics[width=1.0\textwidth]{ 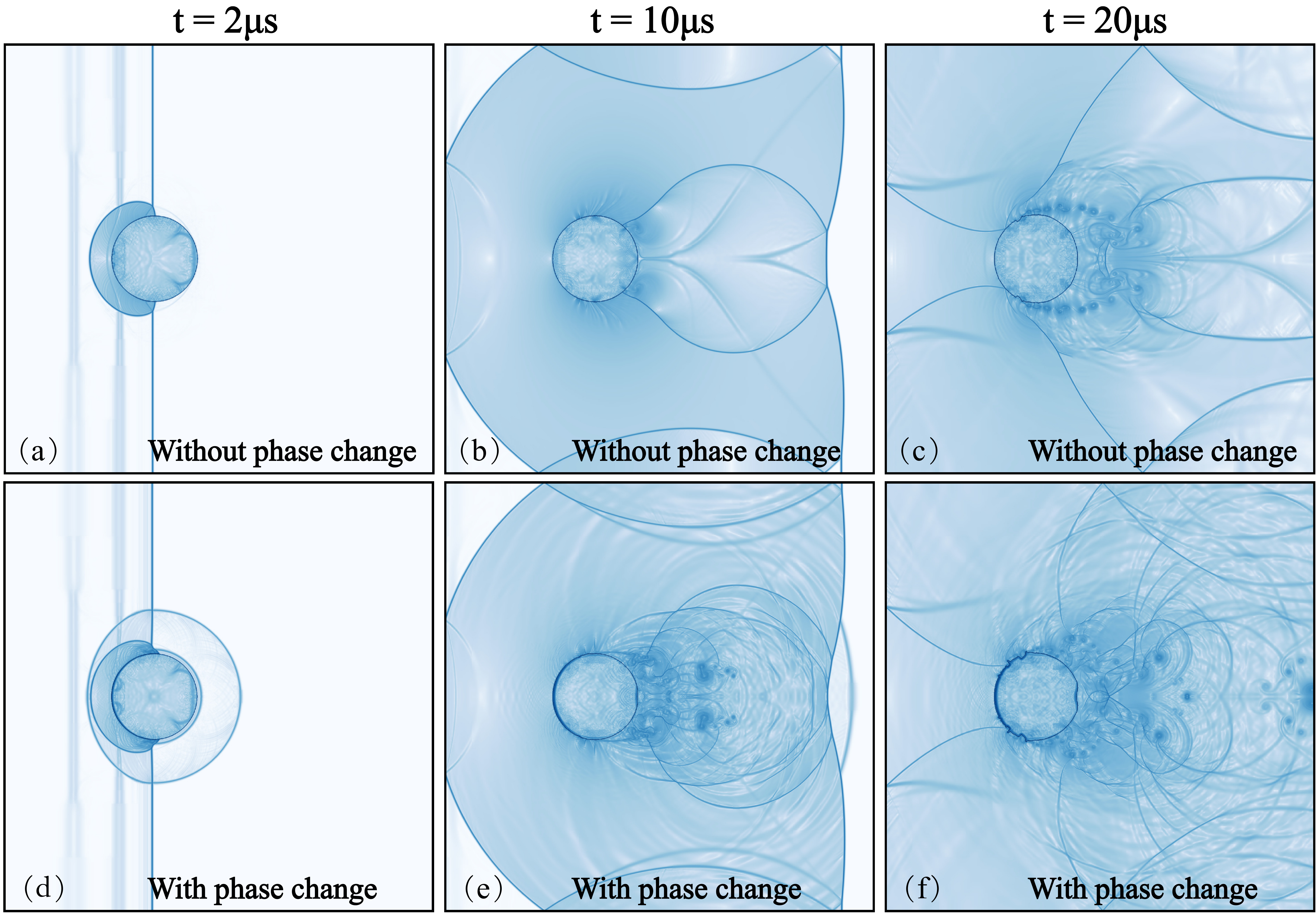}
    \caption{Two-dimensional shock and water droplet interaction: the numerical Schlieren images calculated by $\log{(|\nabla \rho| +1)}$. Each column shows the results at the same time instant with and without phase change. Assuming that the flow is symmetric about the x-axis, only the top half domain is computed with an effective resolution of $2048\times1024$.}
    \label{Fig:shock_drop_water_1}
\end{figure}
\begin{figure}[htbp]
    \centering
    \includegraphics[width=1.0\textwidth]{ 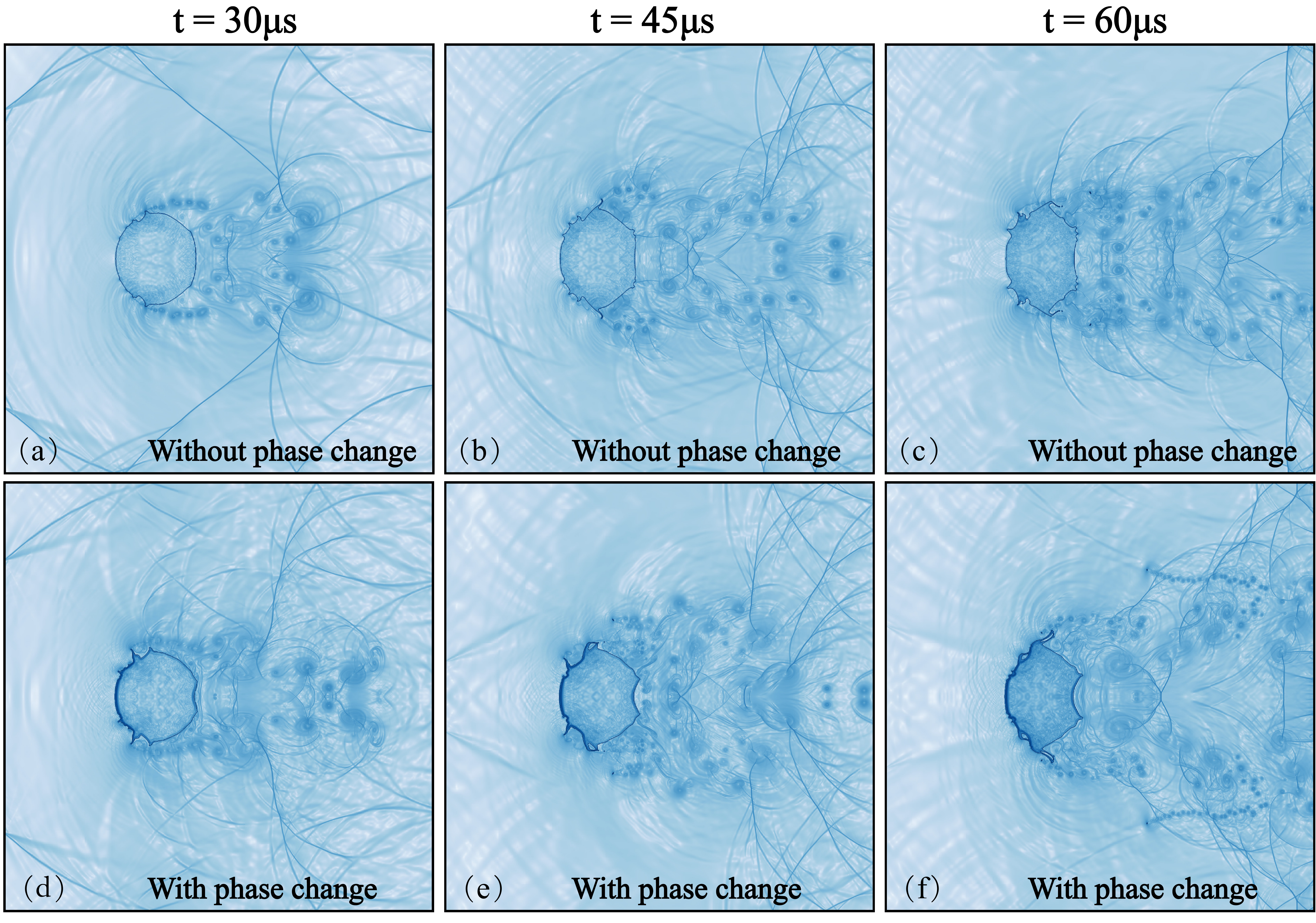}
    \caption{Two-dimensional shock and water droplet interaction: the numerical Schlieren images calculated by $\log{(|\nabla \rho| +1)}$. Each column shows the results at the same time instant with and without phase change. Assuming that the flow is symmetric about the x-axis, only the top half domain is computed with an effective resolution of $2048\times1024$.}
    \label{Fig:shock_drop_water_2}
\end{figure}
Here we study the interaction between a shock wave of Mach number $1.5$ and a cylindrical water droplet of radius $R = 1\ \rm {mm}$. Assuming that the flow is symmetric about the x-axis, the computational domain is the top half of a $10\ \rm{mm}\times10\ \rm{mm}$ square. Initially the shock wave is placed at $x = 2\ \rm{mm}$, and the initial conditions are 
\begin{equation}
    \left\{
    \begin{aligned}
        \rho & =0.57\ \rm{kg/m^3}, u=0\ \rm{m/s},v=0\ \rm{m/s},   \quad \text{pre-shocked vapor},  \\
        p&=1.0\times10^{5}\ \rm{Pa}, T=380.0\ \rm{K}\\
        \rho & =1.09\ \rm{kg/m^3}, u=345.47\ \rm{m/s},v=0\ \rm{m/s},   \quad \text{post-shocked vapor},  \\
        p&=242704\ \rm{Pa}, T=482.47\ \rm{K}\\
        \rho & =1093.67\ \rm{kg/m^3}, u=0\ \rm{m/s},v=0\ \rm{m/s},   \quad \text{water droplet},  \\
        p&=1.0\times10^{5}\ \rm{Pa}, T=373\ \rm{K}\\
        \phi & =-1.0+\sqrt{(x-0.35)^{2}+y^{2}} \ \rm{mm}        \quad level\ set.
    \end{aligned}
    \right.
\end{equation}
Two simulations with and without phase change are conducted with an effective resolution of $2048\times1024$, see Figs. \ref{Fig:shock_drop_water_1} and \ref{Fig:shock_drop_water_2}. When phase change is considered, the model coefficients are set as $\lambda_{evap} = 1.0$ and $\lambda_{cond} = 0.6$ so that the droplet initially evaporates in the pre-shocked vapor. Similarly, the results obtained with phase change exhibit more complex flow structures and larger interface deformations. In compare with the case in Section. \ref{subsubsec4.4.1}, the shock waves in this case travel much faster as the sound speed of water is larger while the interface deformations are smaller due to the stiffness of water. 

\subsubsection{Three-dimensional shock and water droplet interaction}
\label{subsubsec4.4.3}
\begin{figure}[htbp]
    \centering
    \includegraphics[width=1.0\textwidth]{ 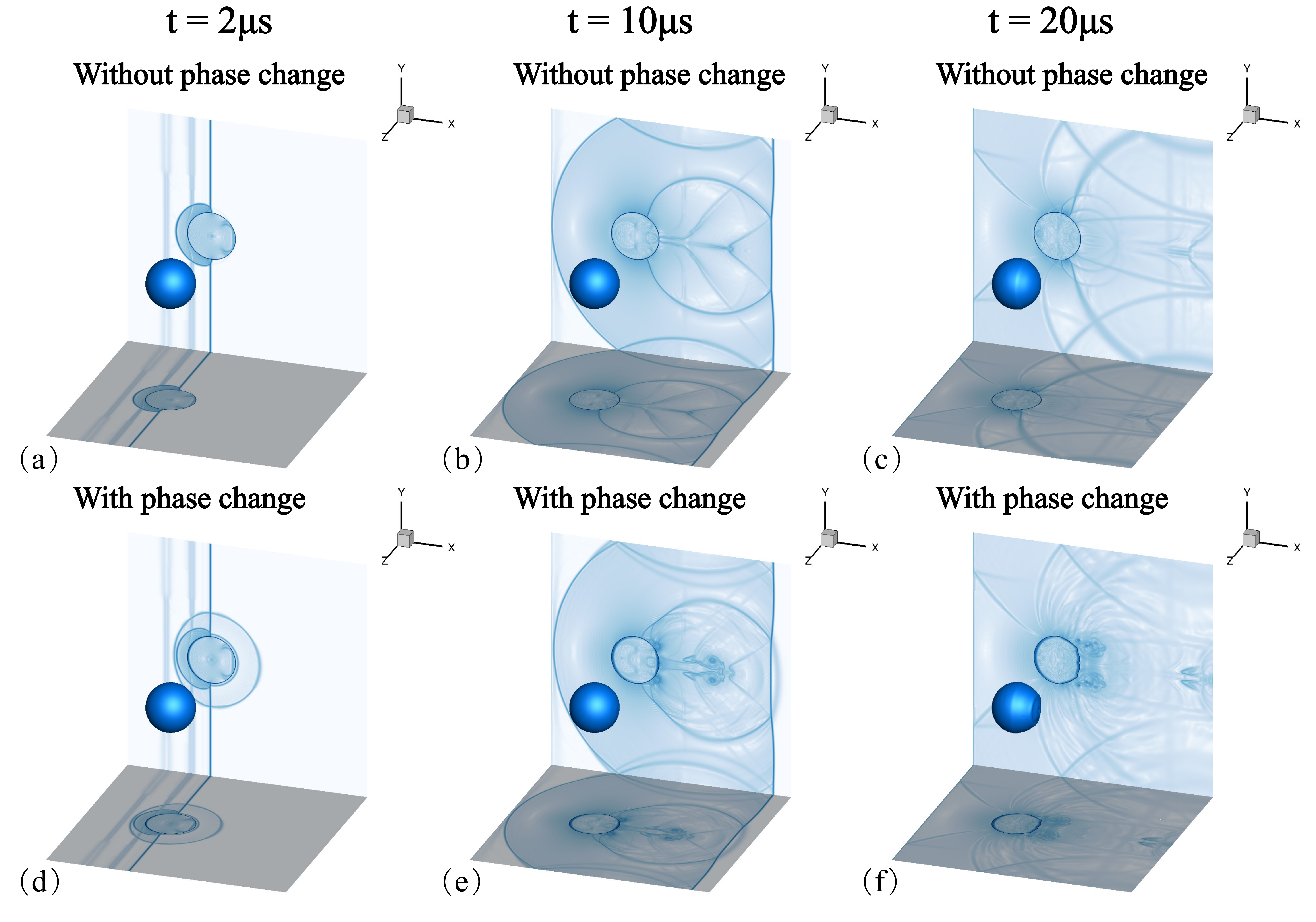}
    \caption{Three-dimensional shock and water droplet interaction. Obtained at the planes of $y = 0$ and $z = 0$, the numerical Schlieren images calculated by $\log{(|\nabla \rho| +1)}$ are plotted, as well as and the droplet interface. Each column shows the results at the same time instant with and without phase change. The flow is assumed to be symmetric in the y-direction and z-direction so that only $1/8$ of the physical domain is computed with an effective resolution of $1024\times512\times512$.}
    \label{Fig:shock_drop_water_3d_1}
\end{figure}
\begin{figure}[htbp]
    \centering
    \includegraphics[width=1.0\textwidth]{ 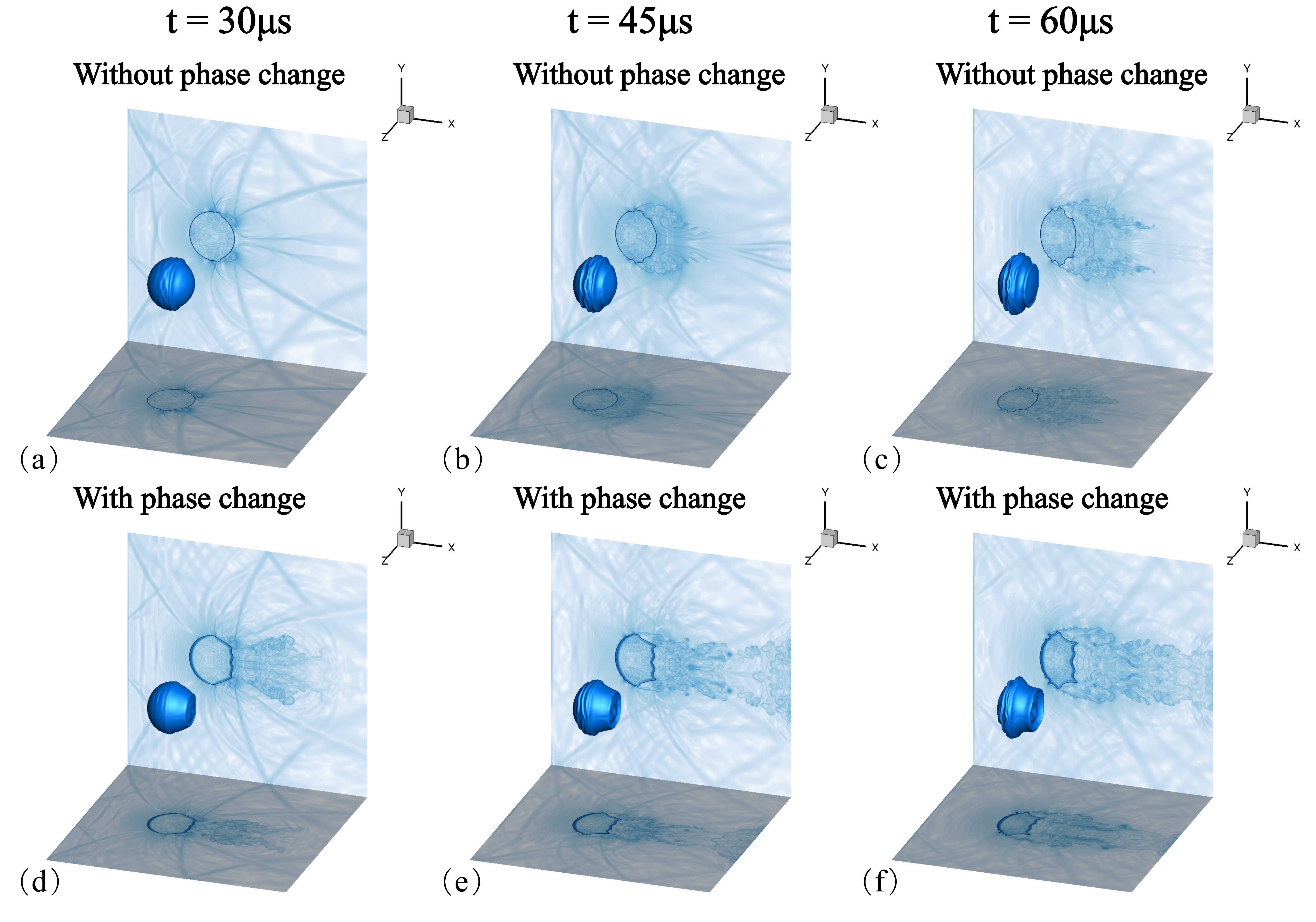}
    \caption{Three-dimensional shock and water droplet interaction. Obtained at the planes of $y = 0$ and $z = 0$, the numerical Schlieren images calculated by $\log{(|\nabla \rho| +1)}$ are plotted, as well as and the droplet interface. Each column shows the results at the same time instant with and without phase change. The flow is assumed to be symmetric in the y-direction and z-direction so that only $1/8$ of the physical domain is computed with an effective resolution of $1024\times512\times512$.}
    \label{Fig:shock_drop_water_3d_2}
\end{figure}
This case is a three-dimensional extension of the case in Section. \ref{subsubsec4.4.2}. All the initial conditions are the same except the interface geometry. Here we consider a three-dimensional spherical water droplet, which is initialized by
\begin{equation}
    \phi =-1.0+\sqrt{(x-0.35)^{2}+y^{2} +y^{2}} \ \rm{mm}.
\end{equation}
As a symmetric configuration, the computational domain is $1/8$ of the physical domain with $10$ mm in the x-direction, $5$ mm in the y-direction and $5$ mm in the z-direction, which is discretized by a MR grid with an effective resolution of $1024\times512\times512$. The numerical results at different time instant are plotted in Figs. \ref{Fig:shock_drop_water_3d_1} and \ref{Fig:shock_drop_water_3d_2}, which confirm that the present method can correctly capture the characteristics of three-dimensional shock-droplet interactions with and without the phase change effects. Due to the three-dimensional effects, the results here are different with those in Figs. \ref{Fig:shock_drop_water_1} and \ref{Fig:shock_drop_water_2}. With a higher resolution, our results show much more details than the previous simulation of Das and Udaykumar \cite{das2020sharp}.
\section{Concluding remarks}
\label{sec5}
In this paper, we have developed a fully conservative sharp-interface method for compressible multiphase flows with phase change. In the previous researches \cite{lee2017phasechange,houim2013ghost,das2020sharp,fechter2017sharp,fechter2018approximate}, the coupling between the liquid phase and the vapor phase is usually implemented through the GFM, which is inherently non-conservative. Based on the work of Hu et al. \cite{hu2006conservative}, the present method hire the interfacial fluxes to ensure strict conservation, which are obtained by solving a general two-phase Riemann problem with phase change. In compare with the methods in Refs. \cite{lauer2012numerical,paula2019analysis}, the present method makes no assumption for the time-scale feature of the phase change process and can guarantee the thermodynamic consistency. A novel approximate Riemann solver employing a four-wave model is developed to solve the Riemann problem efficiently. The original eight-dimensional root-finding procedure in the exact Riemann solver \cite{fechter2017sharp} is simplified as only an iteration of the mass flux. Unlike the approximate Riemann solver of Fechter et al. \cite{fechter2018approximate}, where the energy coupling in the vapor phase is dropped, the present approximate solver impose the jump conditions of all waves strictly. Different strategies for choosing the liquid and the vapor states used in phase change model are discussed in this paper. We have shown that, no matter which model is considered, the adjacent states of the phase interface should be used to ensure numerical consistency. To the author's knowledge, it has not been reported before in the open literature. A number of numerical examples are computed to validate the present method. With good agreements, the results of the present method are compared with the analytical solutions or the results in the previous researches. The influences of viscosity and heat conduction at the two-phase interface will be considered in the future work.

\section{Acknowledgements}
This work was supported by the National Natural Science Foundation of China (No. 11902271), the Fundamental Research Funds for the Central Universities of China (No. G2019KY05102), and the Foundation of National Key Laboratory (No. 6142201190303).

\appendix
\setcounter{equation}{0}
\section{The reduced system of the approximate Riemann solution}
\label{appendix}
The approximate Rieamnn solution leads to a nonlinear system of fifteen dimensions, which can be greatly reduced by using all conditions except the phase change model Eq. \eqref{Eq:phase_change_model}. The unknown variables $\bm{W_M}^*$, $\bm{W_R}^*$, $S_p$, and $S_c$ can be expressed in terms of $j$ so that only a one-dimensional problem needs to be solved, see Eq. \eqref{Eq:reduced_system} for a general description. Here we give the specific expressions of $\bm{W_M}^*$, $\bm{W_R}^*$, $S_p$, and $S_c$. For the sake of brevity, some intermediate variables are defined as
\begin{equation}
    \begin{aligned}
        k &= \rho_L / \rho_R, \\
        l_1 &= \rho_L(V_L - S_L),\quad l_2 = \rho_L(V_L - S_L)V_L + p_L, \\
        l_3 &= \rho_L(V_L - S_L)(e_L+\frac{1}{2}V_L^2) + p_LV_L,\\
        r_1 &= \rho_R(V_R - S_R),\quad r_2 = \rho_R(V_R - S_R)V_R + p_R, \\
        r_3 &= \rho_R(V_R - S_R)(e_R+\frac{1}{2}V_R^2) + p_RV_R.
    \end{aligned}
    \label{Eq:intermediate_variables}
\end{equation}
Then according to the direction of phase change,  $\bm{W_M}^*$, $\bm{W_R}^*$, $S_p$, and $S_c$ can be calculated by
\begin{equation}
    \begin{aligned}
        S_p &= \left\{
        \begin{aligned}
        &\frac{ (k - 1)S_L j^2 + (l_1 S_L - k r_1 S_L - l_2 +r_2 + \sigma \kappa) j + (l_2 - r_2 - \sigma \kappa) l_1}{ (k - 1) j^2  - (k - 1)j r_1 + (l_1 - r_1) l_1   }\quad  &\text{if $j > 0$,}\\
        &\frac{ (k - 1) S_R j^2  + \left[ l_1 S_R - k r_1 S_R + k ( - l_2 + r_2 + \sigma \kappa )\right] j + k (l_2 - r_2 - \sigma \kappa) r_1 }{(k-1) j^2  - (k-1)j l_1 + k(l_1 - r_1) r_1} \quad  &\text{if $j < 0$,}\\
        \end{aligned}
        \right. \\
        S_c & = V_M^* = \left\{
        \begin{aligned}
        & \frac{ j S_L - l_1 S_P + l_2 - r_2 - \sigma \kappa }{ j - r_1 }\quad  &\text{if $j > 0$,} \\
        & \frac{ j S_R - r_1 S_P - l_2 +r_2 + \sigma \kappa  }{ j - l_1}  \quad  &\text{if $j < 0$,}
        \end{aligned}
        \right.\\
        V_L^* & = \left\{
        \begin{aligned}
        & \frac{ j S_L - l_1 S_P }{ j - l_1 } \quad  &\text{if $j > 0$,} \\
        & V_M^*  \quad  &\text{if $j < 0$,}
        \end{aligned}
        \right.
        \quad V_R^* = \left\{
        \begin{aligned}
        & V_M^* \quad  &\text{if $j > 0$,} \\
        & \frac{ j S_R - r_1 S_P }{ j - r_1 }   \quad  &\text{if $j < 0$,}
        \end{aligned}
        \right.\\
        \rho_L^* &= l_1/(V_L^* - S_L),\ \ \ \ p_L^* = l_2 - l_1 V_L^*,\ \ \ \  e_L^* = (l_3 - p_L^* V_L^*)/l_1 - 0.5{V_L^*}^2, \\
        \rho_R^* &= r_1/(V_R^* - S_R),\ \ \ \ p_R^* = r_2 - r_1 V_R^*,\ \ \    e_R^* = (r_3 - p_R^* V_R^*)/r_1 - 0.5{V_R^*}^2, \\
        \rho_M^* &= j/(V_M^* - S_P)\ (j \neq 0), \quad
        p_M^* = \left\{
        \begin{aligned}
        & jV_L^* + p_L^* - j V_M^* - \sigma \kappa  \quad  &\text{if $j > 0$,} \\
        & jV_R^* + p_R^* - j V_M^* + \sigma \kappa  \quad  &\text{if $j < 0$,}
        \end{aligned}
        \right.\\
        e_M^* &= \left\{
        \begin{aligned}
        & e_L^* + \frac{1}{2}{V_L^*}^2 + Q_{lat} + (p_L^* V_L^* - p_M^* V_M^*  - \sigma \kappa S_p)/j - \frac{1}{2}{V_M^*}^2   \quad  &\text{if $j > 0$,} \\
        & e_R^* + \frac{1}{2}{V_R^*}^2 - Q_{lat} + (p_R^* V_R^* - p_M^* V_M^*  + \sigma \kappa S_p)/j - \frac{1}{2}{V_M^*}^2  \quad  &\text{if $j < 0$.}
        \end{aligned}
        \right.\\
    \end{aligned}
\end{equation}
When phase change vanishes, namely $j \to 0$, by using some simple algebra, we can obtain 
\begin{equation}
    \lim_{j\to0^+} S_p = \lim_{j\to0^-} S_p = \lim_{j\to0^+}  S_c = \lim_{j\to0^-} S_c = \frac{l_2 - r_2 - \sigma \kappa}{l_1 - r_1},
    \label{Eq:limit}
\end{equation}
which indicates that the phase interface will coincide with the contact wave. For this limit case, the interfacial fluxes are calculated by using $\bm{W_M}^*$ and $\bm{W_R}^*$.






\bibliographystyle{unsrt}
\bibliography{ref}
\end{document}